\documentclass[12pt]{article}
\newcommand{\bq}{\begin{equation}}
\newcommand{\eq}{\end{equation}}
\newcommand{\ba}{\begin{eqnarray}}
\newcommand{\ea}{\end{eqnarray}}

\newcommand{\nl }{ \nonumber  }

\newcommand{\p}{\partial}
\newcommand{\h}{\hspace{.5cm}}

\newcommand{\la}{\lambda}

\newcommand{\Di}{\left(\p_0-\la^{i}\p_i\right)}
\newcommand{\Dj}{\left(\p_0-\la^{j}\p_j\right)}

\begin{document}
\vspace*{.5cm}
\begin{center}
{\bf ROTATING STRINGS AND D2-BRANES IN TYPE IIA REDUCTION OF
M-THEORY ON $G_2$ MANIFOLD AND THEIR SEMICLASSICAL LIMITS
\vspace*{0.5cm}\\ P. Bozhilov}
\\ {\it Institute for Nuclear Research and Nuclear Energy, \\
Bulgarian Academy of Sciences, \\ 1784 Sofia, Bulgaria\\
E-mail:} bozhilov@inrne.bas.bg
\end{center}
\vspace*{0.5cm}

We consider rotating strings and D2-branes on type IIA background,
which arises as dimensional reduction of M-theory on manifold of
$G_2$ holonomy, dual to $\mathcal{N}=1$ gauge theory in four
dimensions. We obtain exact solutions and explicit expressions for
the conserved charges. By taking the semiclassical limit, we show
that the rotating strings can reproduce only one type of
semiclassical behavior, exhibited by rotating M2-branes on $G_2$
manifolds. Our further investigation leads to the conclusion that
the rotating D2-branes reproduce two types of the semiclassical
energy-charge relations known for membranes in eleven dimensions.


\vspace*{.5cm} {\bf Keywords:} Rotating strings, Rotating
D-branes, String/Gauge Theory Correspondence.


\section{Introduction}
In the recent years, an essential progress has been achieved in
understanding the semiclassical limit of the string/gauge theory
duality \cite{4}. This initiated also an interest in the
investigation of the M-theory lift of this semiclassical
correspondence and in particular, in obtaining new membrane
solutions in curved space-times and finding relations between
their energy and the other conserved charges
\cite{6}-\cite{B0306}. So far, such relations have been obtained
for the following target spaces: $AdS_p\times S^q$ \cite{6},
\cite{10}, \cite{27}, \cite{BRR04}-\cite{0508}, $AdS_4\times
Q^{1,1,1}$ \cite{27}, warped $AdS_5\times M^6$ \cite{27},
11-dimensional $AdS$-black hole \cite{27}, and manifolds of $G_2$
holonomy \cite{27}, \cite{B0306}. In \cite{27}, various rotating
membrane configurations on different $G_2$ holonomy backgrounds
have been studied systematically. In the semiclassical limit
(large conserved charges), the following relations between the
energy and the corresponding charge $K$ have been obtained: $E\sim
K^{1/2}$, $E\sim K^{2/3}$, $E-K\sim K^{1/3}$, $E-K\sim \ln K$. In
\cite{B0306}, rotating membranes on a manifold with exactly known
metric of $G_2$ holonomy \cite{NPB611} have been considered. The
above energy-charge relations, except the last one, have been
reproduced and generalized for the case of more than one conserved
charges. Moreover, examples of more complicated dependence of the
energy on the charges have been found. The most general cases
considered, lead to algebraic equations of third or even forth
order for the $E^2$ as function of up to five conserved momenta.

It seems to us that an interesting task is to check if rotating
strings in type IIA theory in ten dimensions, can reproduce the
energy-charge relations obtained in \cite{27} and \cite{B0306} for
rotating M2-branes.

In this paper, we consider rotating strings on type IIA
background, which arises as dimensional reduction of M-theory on
the manifold of $G_2$ holonomy, discovered in \cite{NPB611}. By
taking the semiclassical limit, we obtain that the rotating
strings can reproduce only one type of semiclassical behavior,
exhibited by rotating M2-branes on $G_2$ manifolds. Namely, $E\sim
K^{1/2}$ and generalizations thereof. Our further investigation
shows that the rotating D2-branes reproduce two types of the
semiclassical energy-charge relations known for membranes in
M-theory. These are generalizations of the dependencies $E\sim
K^{1/2}$ and $E\sim K^{2/3}$.

The paper is organized as follows. In section 2, we describe the
type IIA background, which we will use. In section 3, we settle
the framework, which we will work in. In section 4, we obtain
three types of rotating string solutions and explicit expressions
for the corresponding conserved charges. Then, we take the
semiclassical limit and derive different energy-charge relations.
In section 5, the same is done for rotating D2-branes. Section 6
is devoted to our concluding remarks.

\setcounter{equation}{0}
\section{The type IIA background}
The type IIA background, in which we will search for rotating
string and D2-brane solutions, has the form \cite{NPB611} \ba\nl
ds_{10}^2 &=& r_0^{1/2}C \left\{-(dx^0)^2 + \delta_{IJ}dx^I dx^J +
A^2 \left[ (g^1)^2 + (g^2)^2 \right]\right.
\\ \nl &+& \left. B^2 \left[ (g^3)^2
+ (g^4)^2 \right] + D^2 (g^5)^2 \right\} +
r_0^{1/2}\frac{dr^2}{C},\h (I,J=1,2,3),\h r_0=const,
\\ \label{10db} &&e^\Phi = r_0^{3/4}C^{3/2},\h
F_2 = \sin \theta_1 d\phi_1 \wedge d\theta_1 - \sin \theta_2
d\phi_2 \wedge d\theta_2.\ea Here, $g^1$,...,$g^5$ are given by
\ba\nl
&&g^1=-\sin\theta_1d\phi_1-\cos\psi_1\sin\theta_2d\phi_2+\sin\psi_1
d\theta_2,\\ \nl &&g^2=d\theta_1-\sin\psi_1\sin\theta_2
d\phi_2-\cos\psi_1 d\theta_2,\\ \nl
&&g^3=-\sin\theta_1d\phi_1+\cos\psi_1\sin\theta_2d\phi_2-\sin\psi_1
d\theta_2,\\ \nl &&g^4=d\theta_1+\sin\psi_1\sin\theta_2
d\phi_2+\cos\psi_1 d\theta_2,\\ \nl &&g^5=d\psi_1+\cos\theta_1
d\phi_1+\cos\theta_2 d\phi_2,\ea and the functions $A$, $B$, $C$
and $D$ depend on the radial coordinate $r$ only: \ba\nl
A&=&\frac{1}{\sqrt{12}} \sqrt{(r - 3 r_0/2)(r + 9 r_0/2)},\h
B=\frac{1}{\sqrt{12}} \sqrt{(r + 3 r_0/2)(r - 9 r_0/2)},
\\ \label{G2-3} C&=&\sqrt{\frac{(r - 9 r_0/2)(r + 9 r_0/2)}{(r - 3 r_0/2)(r + 3 r_0/2)}},\h D=r/3.\ea
In (\ref{10db}), $\Phi$ and $F_2$ are the Type IIA dilaton and the
field strength of the Ramond-Ramond one-form gauge field
respectively.

The above ten dimensional background arises as dimensional
reduction of the following solution of the eleven dimensional
supergravity \cite{NPB611} \ba\label{G2-g}
&&l_{11}^{-2}ds_{11}^{2}=-(dx^0)^2 + \delta_{IJ}dx^I dx^J +
ds_{7}^{2},
\\ \nl &&ds^2_7=dr^2/C^2+A^2\left[(g^1)^2+(g^2)^2\right]+B^2\left[(g^3)^2+(g^4)^2\right]+D^2(g^5)^2+r_0\
C^2 (g^6)^2,\ea where $l_{11}$ is the eleven dimensional Planck
length and \ba\nl g^6=d\psi_2+\cos \theta_1 d\phi_1-\cos\theta_2
d\phi_2.\ea

The type IIA solution (\ref{10db}) describes a D6-brane wrapping
the ${\bf S}^3$ in the deformed conifold geometry. For
$r\to\infty$, the metric becomes that of a singular conifold, the
dilaton is constant, and the flux is through the ${\bf S}^2$
surrounding the wrapped D6-brane. For $r - 9r_0/2 = \epsilon \to
0$, the string coupling $e^\Phi$ goes to zero like
$\epsilon^{3/4}$, whereas the curvature blows up as $\epsilon^{-
{3/2}}$ just like in the near horizon region of a flat D6-brane.
This means that classical supergravity is valid for sufficiently
large radius. However, the singularity in the interior is the same
as the one of flat D6 branes, as expected. On the other hand, the
dilaton continuously decreases from a finite value at infinity to
zero, so that for small $r_0$ classical string theory is valid
everywhere. As explained in \cite{NPB611}, the global geometry is
that of a warped product of flat Minkowski space and a non-compact
space, $Y_6$, which for large radius is simply the conifold since
the backreaction of the wrapped D6 brane becomes less and less
important. However, in the interior, the backreaction induces
changes on $Y_6$ away from the conifold geometry. For $r \to 9
r_0/2$, the  ${\bf S}^2$ shrinks to zero size, whereas an ${\bf
S}^3$ of finite size remains. This behavior is similar to that of
the deformed conifold but the two metrics are different.

\setcounter{equation}{0}
\section{The set-up}
The ten dimensional background, described in the previous section,
does not depend on part of the target space coordinates $x^M$,
$M=0,1,...,9$. We denote them by $x^\mu$ and the remaining ones by
$x^a$: $x^M=(x^\mu, x^a)$. Further on, we will use the following
ansatz for the string and D2-brane embedding coordinates
$x^M=X^M(\xi^m)$  \ba\label{sLA} X^\mu(\xi^m)=\Lambda^\mu_m \xi^m
,\h X^a(\xi^m)=Z^a(\xi^p),\h \xi^m=(\xi^0,...,\xi^p), \ea where
$\Lambda^\mu_m$ are constants, $\xi^p=\xi^1$ for the string and
$\xi^p=\xi^2$ for the D2-brane.

\subsection{Rotating strings}
In our further considerations, we will use the Polyakov action for
strings embedded in curved space-time with metric tensor
$g_{MN}(x)$, interacting with a background 2-form gauge field
$b_{MN}(x)$ via Wess-Zumino term \ba\label{pa} &&S^{P}=
=-\frac{T}{2}\int
d^{2}\xi\left(\sqrt{-\gamma}\gamma^{mn}G_{mn}-\varepsilon^{mn}
B_{mn}\right),\\ \nl && \xi^m=(\xi^0,\xi^1),\h m,n = (0,1),\ea
where  \ba\nl G_{mn}= \p_m X^M\p_n X^N g_{MN},\h
B_{mn}=\p_{m}X^{M}\p_{n}X^{N} b_{MN}, \h (\p_m=\p/\p\xi^m),\ea are
the fields induced on the string worldsheet, $\gamma$ is the
determinant of the auxiliary worldsheet metric $\gamma_{mn}$,
$\gamma^{mn}$ is its inverse, and $T=1/2\pi\alpha'$ is the string
tension.

For our background (\ref{10db}), the action (\ref{pa}) reduces to
\ba\label{opa} S^{P}=\int d^{2}\xi\mathcal{L}^P,\h \mathcal{L}^P
=-\frac{T}{2}\sqrt{-\gamma}\gamma^{mn}G_{mn}.\ea The equations of
motion for $X^M$ following from (\ref{opa}) are: \ba \label{sem}
&&-g_{LK}\left[\p_m\left(\sqrt{-\gamma}\gamma^{mn}\p_nX^K\right) +
\sqrt{-\gamma}\gamma^{mn}\Gamma^K_{MN}\p_m X^M\p_n
X^N\right]=0,\ea where \ba\nl
\Gamma_{L,MN}=g_{LK}\Gamma^K_{MN}=\frac{1}{2}\left(\p_Mg_{NL}
+\p_Ng_{ML}-\p_Lg_{MN}\right),\ea are the components of the
symmetric connection corresponding to the metric $g_{MN}$. The
constraints are obtained by varying the action (\ref{opa}) with
respect to $\gamma_{mn}$: \ba\label{oc}
\delta_{\gamma_{mn}}S^P=0\Rightarrow
\left(\gamma^{kl}\gamma^{mn}-2\gamma^{km}\gamma^{ln}\right)G_{mn}=0.\ea

Further on, we will work in conformal gauge
$\gamma^{mn}=\eta^{mn}=diag(-1,1)$, in which the equations of
motion (\ref{sem}) and constraints (\ref{oc}) simplify to \ba
\label{em} &&g_{LK}\eta^{mn}\left(\p_m\p_nX^K + \Gamma^K_{MN}\p_m
X^M\p_n X^N\right)=0.\ea \ba\label{0s} &&G_{00}+G_{11}=0,\\
\label{01s} &&G_{01}=0 .\ea

Taking into account the ansatz (\ref{sLA}), one obtains that the
metric induced on the string worldsheet is given by (the prime is
used for $d/d\xi^1$) \ba\nl &&G_{00}=\Lambda^\mu_0\Lambda^\nu_0
g_{\mu\nu}, \h G_{11}=g_{ab}Z'^a Z'^b + 2\Lambda^\mu_1 g_{\mu
a}Z'^a + \Lambda^\mu_1\Lambda^\nu_1 g_{\mu\nu},\\ \nl &&
G_{01}=\Lambda^\mu_0\left(g_{\mu a}Z'^a + \Lambda^\nu_1
g_{\mu\nu}\right).\ea The Lagrangian density in the action
(\ref{opa}) reduces to \ba\label{La}\mathcal{L}^{A}_s(\xi^1) =
-\frac{T}{2}\left(g_{ab}Z'^aZ'^b+ 2\Lambda_1^\mu g_{\mu a} Z'^a +
\eta^{mn}\Lambda_m^\mu \Lambda_n^\nu g_{\mu \nu}\right).\ea
$\mathcal{L}^{A}_s$ does not depend on $X^\mu$, so the conjugated
momenta \ba\label{cmoms} P_\mu = T\Lambda^\nu_0\int d\xi^1
g_{\mu\nu}\ea are conserved, i.e. they do not depend on the proper
time $\xi^0$.

Let us introduce the density \ba\label{pM}\mathcal{P}_M
\equiv\frac{\p \mathcal{L}^P}{\p (\p_1 X^M)} =-T\sqrt{-
\gamma}\gamma ^{1n}g_{MN}\p_n X^N =-T \left(g_{Mb}Z'^b +
\Lambda_1^\nu g_{M\nu}\right).\ea In terms of $\mathcal{P}_M$, the
equations of motion (\ref{em}) read \ba\label{secm}
\left[\mathcal{P}_\mu(\xi^1)\right]'=0,
\\ \label{srem} \left(\mathcal{P}_a\right)' - \frac{\p\mathcal{L}_s^A}{\p Z^a}=0.\ea
The equations (\ref{secm}) mean that $\mathcal{P}_\mu$ are
constants of the motion: $\mathcal{P}_\mu=constants$. The
remaining equations (\ref{srem}) may be rewritten as
\ba\label{fems} &&g_{ab}Z''^b + \Gamma_{a,bc}Z'^bZ'^c =
\frac{1}{2}\p_a \mathcal{U} + 2\p_{[a}\mathcal{A}_{b]}Z'^b,
\\ \nl &&\p_{[a}\mathcal{A}_{b]}= \frac{1}{2}
\left(\p_a\mathcal{A}_{b} - \p_b\mathcal{A}_{a}\right).\ea In
(\ref{fems}), an effective scalar potential $\mathcal{U}$ and an
effective 1-form gauge field $\mathcal{A}_a$ appeared. They are
given by \ba\nl\mathcal{U}=
\eta^{mn}\Lambda^{\mu}_{m}\Lambda^{\nu}_{n}g_{\mu\nu} +
\frac{2\Lambda^{\mu}_{1}\mathcal{P}_\mu}{T},\h \nl \mathcal{A}_a=
\Lambda_1^\mu g_{a\mu}.\ea

The constraints (\ref{0s}), (\ref{01s}) take the form
\ba\label{ecs} g_{ab}Z'^aZ'^b = \mathcal{U},\h
\Lambda^\mu_0\left(g_{\mu a}Z'^a + \Lambda^\nu_1
g_{\mu\nu}\right)=0.\ea

Here, we are interested in obtaining rotating string solutions for
which the conditions $\mathcal{P}_\mu=constants$ and the second
constraint in (\ref{ecs}) are identically satisfied by appropriate
choice of the embedding parameters $\Lambda^\mu_m$. Then, the
problem reduces to solving the equations of motion (\ref{fems})
and the first constraint in (\ref{ecs}). We further restrict
ourselves to the simplest case, when the embedding is such that
the background seen by the string depends only on the radial
coordinate $r$. In this case, the solution is \cite{32}
\ba\label{1dc}\xi^1\left(r\right)=\int_{r_{min}}^{r}
\left(\frac{g_{rr}}{\mathcal{U}}\right)^{1/2}dt .\ea On the
solution (\ref{1dc}), the conserved generalized momenta
(\ref{cmoms}) take the form \ba\label{scm}P_\mu =
2T\Lambda^\nu_0\int_{r_{min}}^{r_{max}}g_{\mu\nu}
\left(\frac{g_{rr}}{\mathcal{U}}\right)^{1/2}dt .\ea

\subsection{Rotating D2-branes}
The Dirac-Born-Infeld type action for D2-brane in ten dimensional
space-time with metric tensor $g_{MN}(x)$, interacting with a
background 3-form Ramond-Ramond gauge field $c_{MNP}(x)$ via
Wess-Zumino term, can be written in string frame as
\ba\label{dbia} S^{DBI}&=&-T_{D2}\int d^{3}\xi
\Bigl\{e^{-\Phi}\sqrt{-\det\left(G_{mn} + B_{mn} +
2\pi\alpha'F_{mn}\right)}\\ \nl &-&\frac{ \varepsilon^{m_1 m_2
m_3}}{3!} \p_{m_1}X^{M_1} \p_{m_2}X^{M_2}\p_{m_3}X^{M_3} c_{M_1
M_2 M_3}\Bigr\}.\ea Here, $T_{D2}$ is the D2-brane tension,
$G_{mn}$, $B_{mn}$ and $\Phi$ are the pullbacks of the background
metric, antisymmetric tensor and dilaton to the D2-brane
worldvolume, while $F_{mn}$ is the field strength of the
worldvolume $U(1)$ gauge field $A_m$: $F_{mn}=2\p_{[m}A_{n]}$. For
our background, (\ref{dbia}) reduces to \footnote{For $A_m=\p_m
f$.} \ba\nl S^{DBI}&=&-T_{D2}\int d^{3}\xi e^{-\Phi}\sqrt{-\det
G_{mn}},\ea which is classically equivalent to the following
action \cite{NPB} \ba\label{oda} S_{D2}=\int
d^{3}\xi\mathcal{L}_{D2}=\int d^{3}\xi
\frac{e^{-\Phi}}{4\lambda^0}\Bigl[G_{00}-2\lambda^i G_{0i} +
\lambda^i\lambda^j G_{ij} - \left(2\lambda^0T_{D2}\right)^2 \det
G_{ij}\Bigr],\ea where $\lambda^m=(\lambda^0$, $\lambda^i)$,
$(i,j=1,2)$ are Lagrange multipliers, which equations of motion
generate the {\it independent} constraints \ba\label{0} &&
G_{00}-2\lambda^{j}G_{0j}+\lambda^{i}\lambda^{j}G_{ij}
+\left(2\lambda^0T_{D2}\right)^2\det G_{ij}=0,\\
\label{0j} &&G_{0j}-\lambda^{i}G_{ij}=0.\ea Further on, we will
use the action (\ref{oda}) because it does not contain square root
opposite to the DBI type action (\ref{dbia}), thus avoiding the
introduction of additional nonlinearities in the equations of
motion.

The equations of motion for $X^M$ following from (\ref{oda}), in
the worldvolume gauge $\lambda^m=consants$, are
$(\mathbf{G}\equiv\det{G_{ij}})$ \ba\label{eqm}
&&g_{MN}\left[\Di\Dj X^N - \left(2\lambda^0T_{D2}\right)^2
\p_i\left(\mathbf{G}G^{ij}\p_j X^N\right)\right]\\ \nl
&&+\left[\Gamma_{M,NK}-\left(g_{MK}\p_N\Phi-\frac{1}{2}g_{NK}\p_M\Phi\right)\right]\Di
X^N \Dj X^K \\ \nl &&- \left(2\lambda^0T_{D2}\right)^2
\mathbf{G}\left[\left(\Gamma_{M,NK}-g_{MK}\p_N\Phi\right)
G^{ij}\p_i X^N \p_j X^K+\frac{1}{2}\p_M\Phi\right]=0.\ea

In practice, it turns out that using the diagonal gauge
$\lambda^{i}=0$ simplify the considerations a lot \cite{B0705}.
That is why, we restrict ourselves namely to this gauge from now
on. In this case, (\ref{oda}), (\ref{0}), (\ref{0j}) and
(\ref{eqm}) reduce to \ba\label{ogfa} S_{D2}^{gf}=\int
d^{3}\xi\mathcal{L}_{D2}^{gf}=\int d^{3}\xi
\frac{e^{-\Phi}}{4\lambda^0}\Bigl[G_{00} -
\left(2\lambda^0T_{D2}\right)^2\mathbf{G}\Bigr],\ea \ba\label{00}
&&G_{00}+\left(2\lambda^0T_{D2}\right)^2\mathbf{G}=0,\\
\label{rci} &&G_{0i}=0,\ea \ba\label{gfem} &&g_{MN}\left[\p_0^2
X^N - \left(2\lambda^0T_{D2}\right)^2
\p_i\left(\mathbf{G}G^{ij}\p_j X^N\right)\right]\\ \nl
&&+\left[\Gamma_{M,NK}-\left(g_{MK}\p_N\Phi-\frac{1}{2}g_{NK}\p_M\Phi\right)\right]\p_0
X^N \p_0 X^K \\ \nl &&- \left(2\lambda^0T_{D2}\right)^2
\mathbf{G}\left[\left(\Gamma_{M,NK}-g_{MK}\p_N\Phi\right)
G^{ij}\p_i X^N \p_j X^K+\frac{1}{2}\p_M\Phi\right]=0.\ea

Taking into account the ansatz (\ref{sLA}), one obtains that the
metric induced on the D2-brane worldvolume is given by (the prime
is used for $d/d\xi^2$) \ba\nl &&G_{00}=\Lambda^\mu_0\Lambda^\nu_0
g_{\mu\nu},\h G_{11}=\Lambda^\mu_1\Lambda^\nu_1 g_{\mu\nu}, \h
G_{22}=g_{ab}Z'^a Z'^b + 2\Lambda^\mu_2 g_{\mu a}Z'^a +
\Lambda^\mu_2\Lambda^\nu_2 g_{\mu\nu},\\ \nl
&&G_{01}=\Lambda^\mu_0\Lambda^\nu_1 g_{\mu\nu},\h
G_{02}=\Lambda^\mu_0\left(g_{\mu a}Z'^a + \Lambda^\nu_2
g_{\mu\nu}\right),\h G_{12}=\Lambda^\mu_1\left(g_{\mu a}Z'^a +
\Lambda^\nu_2 g_{\mu\nu}\right).\ea Correspondingly, the
Lagrangian density in the action (\ref{ogfa}) reduces to
\ba\label{old} \mathcal{L}^{A}(\xi^2)
=\frac{1}{4\lambda^0}\left(\tilde{K}_{ab}Z'^aZ'^b +
2\tilde{A}_{a}Z'^a - \tilde{V}\right),\ea where \ba\nl
\tilde{K}_{ab}&=&-\left(2\lambda^0T_{D2}\right)^2
\Lambda^\mu_1\Lambda^\nu_1
\left.\right.\left(g_{ab}g_{\mu\nu}-g_{a\mu}g_{b\nu}\right)e^{-\Phi},\\
\nl \tilde{A}_{a}&=&\left(2\lambda^0T_{D2}\right)^2
\Lambda^\mu_1\Lambda^\nu_1\Lambda^\rho_2
\left(g_{a\mu}g_{\nu\rho}-g_{a\rho}g_{\mu\nu}\right)e^{-\Phi},\\
\nl \tilde{V}&=&\left[-\Lambda^\mu_0\Lambda^\nu_0 g_{\mu\nu}+
\left(2\lambda^0T_{D2}\right)^2
\Lambda^\mu_1\Lambda^\nu_1\Lambda^\rho_2\Lambda^\lambda_2
\left(g_{\mu\nu}g_{\rho\lambda}-g_{\mu\rho}g_{\nu\lambda}\right)\right]e^{-\Phi}.\ea
As far as $\mathcal{L}^{A}$ does not depend on $X^\mu$, the
momenta \ba\label{cmom} P_\mu =
\frac{\Lambda^\nu_0}{2\lambda^{0}}\int\int d\xi^1 d\xi^2
g_{\mu\nu}e^{-\Phi}\ea are conserved.

If we introduce the densities \ba\nl
\mathcal{P}^i_M=\frac{\p\mathcal{L}_{D2}^{gf}}{\p\left(\p_i
X^M\right)},\ea the equations of motion (\ref{gfem}) acquire the
form \ba\label{ecm} \left[\mathcal{P}^2_\mu(\xi^2)\right]'=0,
\\ \label{rem} \left(\mathcal{P}^2_a\right)' - \frac{\p\mathcal{L}^A}{\p Z^a}=0.\ea
The equations (\ref{ecm}) just state that $\mathcal{P}^2_\mu$ are
constants of the motion: \ba\label{cm} \mathcal{P}^2_\mu =
2\lambda^0T_{D2}^2e^{-\Phi}\Lambda^\nu_1\Lambda^\rho_1\left[\left(g_{\mu\nu}
g_{\rho a}-g_{\mu a}g_{\nu\rho}\right)Z'^a +\Lambda^\lambda_2
\left(g_{\mu\nu}g_{\rho\lambda}-g_{\mu\lambda}g_{\nu\rho}\right)\right]=constants.\ea
In the case under consideration, this is possible only for
$\mathcal{P}^2_\mu = 0$. The remaining equations (\ref{rem}) may
be rewritten as \ba\label{rema} \tilde{K}_{ab}Z''^b +
\tilde{\Gamma}_{a,bc}Z'^b Z'^c - 2\p_{[a}\tilde{A}_{b]}Z'^b +
\frac{1}{2}\p_a \tilde{V} = 0,\ea where \ba\nl
\tilde{\Gamma}_{a,bc}= \frac{1}{2}\left(\p_b \tilde{K}_{ca}+\p_c
\tilde{K}_{ba}-\p_a \tilde{K}_{bc}\right).\ea

The constraints (\ref{00}) and (\ref{rci}) take the form
\ba\label{00e} &&\tilde{K}_{ab}Z'^aZ'^b + \tilde{V}=0,
\\ \label{01} &&\Lambda^\mu_0\Lambda^\nu_1 g_{\mu\nu}=0,
\\ \label{02} &&\Lambda^\mu_0\left(g_{\mu a}Z'^a + \Lambda^\nu_2
g_{\mu\nu}\right)=0.\ea

We will search for D2-brane solutions for which the conditions
(\ref{01}), (\ref{02}) and $\mathcal{P}^2_\mu=0$ are identically
satisfied due to appropriate choice of the embedding parameters
$\Lambda^\mu_m$. Then, the investigation of the D2-brane dynamics
reduces to the problem of solving the equations of motion
(\ref{rema}) and the remaining constraint (\ref{00e}). In this
article, we restrict ourselves to the simplest case, when the
embedding is such that the background seen by the D2-brane depends
on the radial coordinate $r$ only. Then, the constraint
(\ref{00e}) is first integral of the equation of motion
(\ref{rema}) for $Z^a(\xi^2)=r(\xi^2)$, and the solution is given
by \ba\label{D1dc}\xi^2\left(r\right)=\int_{r_{min}}^{r}
\left(-\frac{\tilde{K}_{rr}}{\tilde{V}}\right)^{1/2}dt.\ea On the
solution (\ref{D1dc}), the conserved generalized momenta
(\ref{cmom}) take the form \ba\label{Dcm}
P_\mu=\frac{\pi\Lambda_0^\nu}{\lambda^0}
\int_{r_{min}}^{r_{max}}g_{\mu\nu}\left(-\frac{\tilde{K}_{rr}}{\tilde{V}}\right)^{1/2}
e^{-\Phi}dt.\ea

\setcounter{equation}{0}
\section{Rotating string solutions, conserved charges and
\\their semiclassical limits}
As we already mentioned in the previous section, we are interested
here in obtaining rotating string solutions, for which the
embedding is such that the background seen by the string depends
only on the radial coordinate $r$. This leads to the following
three cases\footnote{For all of them $F_2=0$.}

\paragraph{1. $\psi_1$, $\phi_1$, $\phi_2$ fixed to $\psi_1^0$,
$\phi_1^0$, $\phi_2^0$} \ba\nl ds^2 &=& r_0^{1/2}\left\{C
\left[-(dx^0)^2 + \delta_{IJ}dx^I dx^J + \left(A^2+B^2\right)
\left(d\theta_1^2+d\theta_2^2\right)\right.\right. \\
\label{1G}
&&-\left.\left.2\left(A^2-B^2\right)\cos\psi_1^0d\theta_1d\theta_2\right]+
\frac{dr^2}{C}\right\}.\ea

\paragraph{2. $\psi_1$, $\phi_1$, $\theta_2$ fixed to $\psi_1^0$,
$\phi_1^0$, $\theta_2^0$} \ba\nl ds^2 &=& r_0^{1/2}\left\{C
\left\{-(dx^0)^2 + \delta_{IJ}dx^I dx^J +
\left(A^2+B^2\right)d\theta_1^2\right.\right.
\\ \label{3G'} &&+\left.\left.\left[\left(A^2+B^2\right)\sin^2\theta_2^0+D^2\cos^2\theta_2^0\right]d\phi_2^2
\right.\right. \\ \nl
&&-\left.\left.2\left(A^2-B^2\right)\sin\psi_1^0\sin\theta_2^0
d\theta_1d\phi_2\right\}+ \frac{dr^2}{C}\right\}.\ea

\paragraph{3. $\psi_1$, $\theta_1$, $\theta_2$ fixed to $\psi_1^0$,
$\theta_1^0$, $\theta_2^0$} \ba\nl ds^2 &=& r_0^{1/2}\left\{C
\left\{-(dx^0)^2 + \delta_{IJ}dx^I dx^J +
\left[\left(A^2+B^2\right)\sin^2\theta_1^0+D^2\cos^2\theta_1^0\right]d\phi_1^2\right.\right.
\\ \label{2G'} &&+\left.\left.\left[\left(A^2+B^2\right)\sin^2\theta_2^0+D^2\cos^2\theta_2^0\right]d\phi_2^2
\right.\right. \\ \nl
&&+\left.\left.2\left[\left(A^2-B^2\right)\cos\psi_1^0\sin\theta_1^0\sin\theta_2^0
+D^2\cos\theta_1^0\cos\theta_2^0\right] d\phi_1d\phi_2\right\}+
\frac{dr^2}{C}\right\}.\ea There are also other possibilities, but
they lead to the same type of metrics with respect to other
coordinates.

Let us begin with considering string moving in the background
(\ref{1G}). In this case, the most general ansatz of the type
(\ref{sLA}), which ensures that the conditions $\mathcal{P}_\mu=0$
and the second constraint in (\ref{ecs}) are identically satisfied
is \ba\label{se1} X^0=\Lambda_0^0\xi^0, \h X^I=\Lambda_0^I\xi^0,
\h r=r(\xi^1),\h \theta_1=\Lambda_0^{\theta_1}\xi^0,\h
\theta_2=\Lambda_0^{\theta_2}\xi^0.\ea It corresponds to string
extended in the radial direction $r$, and rotating in the planes
given by the angles  $\theta_1$ and $\theta_2$ with angular
momenta $P_{\theta_1}$ and $P_{\theta_2}$. At the same time, the
string moves along $x^0$-coordinate with constant energy $E$, and
along $x^I$ with constant momenta $P_{I}$.

From the first constraint in (\ref{ecs}), \ba\nl g_{rr}r'^2 -
\mathcal{U}=\frac{r_0^{1/2}}{C}r'^2-
r_0^{1/2}C\left(v_0^2-\Lambda_-^2A^2-\Lambda_+^2B^2\right)=0,\ea
where \ba\label{vL1} &&v_0^2=
\left(\Lambda_0^0\right)^2-\delta_{IJ}\Lambda_0^I\Lambda_0^J
=\left(\Lambda_0^0\right)^2-\mathbf{\Lambda}_0^2,\\ \nl
&&\Lambda_{\pm}^2=\left(\Lambda_0^{\theta_1}\right)^2 +
\left(\Lambda_0^{\theta_2}\right)^2 \pm
2\Lambda_0^{\theta_1}\Lambda_0^{\theta_2}\cos\psi_1^0,\ea one
obtains the turning points of the effective one-dimensional
periodic motion by solving the equation $r'=0$. In the case under
consideration, the result is\footnote{For all string and D2-brane
solutions we are considering here, $r_{min}=9r_0/2\equiv 3l$.}
\ba\nl &&r_{min}=9r_0/2\equiv 3l,\h
r_{max}=r_1=l\left[2\sqrt{\frac{k^2+3}{4}+\frac{3v_0^2}{l^2\left(\Lambda_+^2+\Lambda_-^2\right)}}+k\right]>3l,
\\ \label{mm} &&r_2=-l\left[2\sqrt{\frac{k^2+3}{4}+
\frac{3v_0^2}{l^2\left(\Lambda_+^2+\Lambda_-^2\right)}}-k\right]<0,
\h k=\frac{\Lambda_+^2-\Lambda_-^2}{\Lambda_+^2+\Lambda_-^2}.\ea

In accordance with (\ref{1dc}), we obtain the following expression
for the string solution ($\Delta r=r-3l$, $\Delta r_1=r_1-3l$)
\ba\label{ssG1} &&\xi^1(r)=
\frac{8}{\left(\Lambda_+^2+\Lambda_-^2\right)^{1/2}}\left[\frac{l\Delta
r}{(3l-r_2)\Delta r_1}\right]^{1/2}\times
\\ \nl &&F_D^{(5)}\left(1/2;-1/2,-1/2,1/2,1/2,1/2;3/2;
-\frac{\Delta r}{2l},-\frac{\Delta r}{4l},-\frac{\Delta
r}{6l},-\frac{\Delta r}{3l-r_2}, \frac{\Delta r}{\Delta
r_1}\right),\ea where $F_D^{(5)}$ is hypergeometric function of
five variables\footnote{The definition and some properties of the
hypergeometric functions
$F_D^{(n)}(a;b_1,\ldots,b_n;c;z_1,\ldots,z_n)$ are given in
Appendix A.}.

Now, we can compute the conserved momenta on the obtained
solution. According to (\ref{scm}), they are ($E=-P_0$):
\ba\label{EP1} \frac{E}{\Lambda_0^0}=\frac{P_I}{\Lambda_0^I}&=&
T\left[\frac{2^7l\Delta r_1}{\left(\Lambda_+^2+
\Lambda_-^2\right)(3l-r_2)}\right]^{1/2}\left(1+\frac{\Delta
r_1}{3l-r_2}\right)^{-1/2}
\\ \nl &&\times F_D^{(1)}\left(1/2;1/2;3/2;\frac{1}{1+\frac{3l-r_2}{\Delta
r_1}}\right),\ea \ba\label{pt12}
&&P_{\theta_1}=\left(\Lambda_0^{\theta_1}-\Lambda_0^{\theta_2}\cos\psi_1^0\right)I_A
+\left(\Lambda_0^{\theta_1}+\Lambda_0^{\theta_2}\cos\psi_1^0\right)I_B,\\
\nl
&&P_{\theta_2}=\left(\Lambda_0^{\theta_2}-\Lambda_0^{\theta_1}\cos\psi_1^0\right)I_A
+\left(\Lambda_0^{\theta_2}+\Lambda_0^{\theta_1}\cos\psi_1^0\right)I_B,\ea
where \ba\label{IA} I_A&=&T\left[\frac{2^7l^5\Delta
r_1}{\left(\Lambda_+^2+
\Lambda_-^2\right)(3l-r_2)}\right]^{1/2}\left(1+\frac{\Delta
r_1}{2l}\right)\left(1+\frac{\Delta r_1}{6l}\right)
\left(1+\frac{\Delta r_1}{3l-r_2}\right)^{-1/2}
\\ \nl &&\times F_D^{(3)}\left(1/2;-1,-1,1/2;3/2;\frac{1}{1+\frac{2l}{\Delta
r_1}},\frac{1}{1+\frac{6l}{\Delta
r_1}},\frac{1}{1+\frac{3l-r_2}{\Delta r_1}}\right),\ea
\ba\label{IB} I_B&=&\frac{T}{9}\left[\frac{2^9\left(l\Delta
r_1\right)^3}{\left(\Lambda_+^2+
\Lambda_-^2\right)(3l-r_2)}\right]^{1/2}\left(1+\frac{\Delta
r_1}{4l}\right) \left(1+\frac{\Delta r_1}{3l-r_2}\right)^{-1/2}
\\ \nl &&\times F_D^{(2)}\left(1/2;-1,1/2;5/2;\frac{1}{1+\frac{4l}{\Delta
r_1}},\frac{1}{1+\frac{3l-r_2}{\Delta r_1}}\right).\ea

Our next task is to find the relation between the energy $E$ and
the other conserved quantities $P_I$, $P_{\theta_1}$,
$P_{\theta_2}$, in the semiclassical limit (large conserved
charges), which corresponds to $r_1\to\infty$. In this limit,
\ba\nl \frac{E}{\Lambda_0^0}=\frac{P_I}{\Lambda_0^I}= \frac{\pi
T\left(2^3l\right)^{1/2}}{\left(\Lambda_+^2+\Lambda_-^2\right)^{1/2}},
\h I_A=I_B=\frac{\pi
T\left(2l\right)^{1/2}v_0^2}{\left(\Lambda_+^2+\Lambda_-^2\right)^{3/2}},
\ea which leads to \ba\label{EG1} E^2=\mathbf{P}^2+2\pi T
\left(6r_0\right)^{1/2}\left(P_{\theta_1}^2+P_{\theta_2}^2\right)^{1/2},\h
\mathbf{P}^2=\delta_{IJ}P_IP_J.\ea This is a generalization of the
energy-charge relation $E\sim K^{1/2}$ for the case $P_I\ne 0$ and
two conserved angular momenta $P_{\theta_1}$, $P_{\theta_2}$.
Thus, the above string configuration has the same semiclassical
behavior as the membrane in (4.20) of \cite{B0306}, which is given
by the relation \ba\nl E^2=\mathbf{P}^2 + 2\sqrt{6}\pi^2 T_{M2}
l_{11}^3\mid\mathbf{\Lambda}_1\mid
\left(P^2_{\theta}+P^2_{\tilde{\theta}}\right)^{1/2}.\ea

Now, let us consider rotating string on the background
(\ref{3G'}). To ensure that the conditions $\mathcal{P}_\mu=0$ and
the second constraint in (\ref{ecs}) are satisfied, we have to
choose the following embedding \ba\label{se2}
X^0=\Lambda_0^0\xi^0, \h X^I=\Lambda_0^I\xi^0, \h r=r(\xi^1),\h
\theta_1=\Lambda_0^{\theta}\xi^0,\h
\phi_2=\Lambda_0^{\phi}\xi^0.\ea This ansatz is analogous to the
previous one, with $\theta_2$ replaced by $\phi_2$. The first
constraint in (\ref{ecs}) now reads, \ba\label{ecG3'} &&g_{rr}r'^2
- \mathcal{U}=\frac{r_0^{1/2}}{C}r'^2-
r_0^{1/2}C\left(v_0^2-\bar{\Lambda}_-^2A^2-\bar{\Lambda}_+^2B^2-\Lambda_D^2D^2\right)=0,
\ea where  $v_0^2$ is given in (\ref{vL1}) and  \ba\nl
&&\bar{\Lambda}_{\pm}^2=\left(\Lambda_0^{\theta}\right)^2 +
\left(\Lambda_0^{\phi}\right)^2\sin^2\theta_2^0 \pm
2\Lambda_0^{\theta}\Lambda_0^{\phi}\sin\psi_1^0\sin\theta_2^0,
\\ \label{LBar} &&\Lambda_D^2=\left(\Lambda_0^{\phi}\right)^2\cos^2\theta_2^0 .\ea
From here, one obtains the solutions of the equation $r'=0$:
\ba\nl &&r_{min}=9r_0/2\equiv 3l,\h r_{max}=r_1>3l,\h r_2<0.\ea
The rotating string solution $\xi^1(r)$ expresses through the same
hypergeometric function as in (\ref{ssG1}), but now depends on
different parameters \ba\label{ssG3'} &&\xi^1(r)=
\frac{8}{\left(\bar{\Lambda}_+^2+\bar{\Lambda}_-^2
+4\Lambda_D^2/3\right)^{1/2}}\left[\frac{l\Delta r}{(3l-r_2)\Delta
r_1}\right]^{1/2}\times
\\ \nl &&F_D^{(5)}\left(1/2;-1/2,-1/2,1/2,1/2,1/2;3/2;
-\frac{\Delta r}{2l},-\frac{\Delta r}{4l},-\frac{\Delta
r}{6l},-\frac{\Delta r}{3l-r_2}, \frac{\Delta r}{\Delta
r_1}\right).\ea The same is true for $E$ and $P_I$ (compare with
(\ref{EP1})) \ba\label{EP2}
\frac{E}{\Lambda_0^0}=\frac{P_I}{\Lambda_0^I}&=&
8T\left[\frac{2l\Delta
r_1}{\left(\bar{\Lambda}_+^2+\bar{\Lambda}_-^2
+4\Lambda_D^2/3\right)(3l-r_2)}\right]^{1/2}\left(1+\frac{\Delta
r_1}{3l-r_2}\right)^{-1/2}
\\ \nl &&\times F_D^{(1)}\left(1/2;1/2;3/2;\frac{1}{1+\frac{3l-r_2}{\Delta
r_1}}\right).\ea For the conserved angular momenta $P_\theta$ and
$P_\phi$, (\ref{scm}) gives \ba\label{am12}
P_{\theta}&=&\left(\Lambda_0^{\theta}-\Lambda_0^{\phi}\sin\psi_1^0\sin\theta_2^0\right)J_A
+\left(\Lambda_0^{\theta}+\Lambda_0^{\phi}\sin\psi_1^0\sin\theta_2^0\right)J_B,
\\ \nl
P_{\phi}&=&\left(\Lambda_0^{\phi}\sin\theta_2^0-\Lambda_0^{\theta}\sin\psi_1^0\right)\sin\theta_2^0J_A
+\left(\Lambda_0^{\phi}\sin\theta_2^0+\Lambda_0^{\theta}\sin\psi_1^0\right)\sin\theta_2^0J_B
\\ \nl &&+\Lambda_0^{\phi}\cos^2\theta_2^0 J_D, \ea where \ba\label{JA} J_A &=&
8T\left[\frac{2l^5\Delta
r_1}{\left(\bar{\Lambda}_+^2+\bar{\Lambda}_-^2
+4\Lambda_D^2/3\right)(3l-r_2)}\right]^{1/2}\left(1+\frac{\Delta
r_1}{2l}\right)\left(1+\frac{\Delta r_1}{6l}\right)
\\ \nl &&\times\left(1+\frac{\Delta r_1}{3l-r_2}\right)^{-1/2}
F_D^{(3)}\left(1/2;-1,-1,1/2;3/2;\frac{1}{1+\frac{2l}{\Delta
r_1}},\frac{1}{1+\frac{6l}{\Delta
r_1}},\frac{1}{1+\frac{3l-r_2}{\Delta r_1}}\right),\ea \ba\nl
J_B&=& \frac{16}{9}T\left[\frac{2(l\Delta
r_1)^3}{\left(\bar{\Lambda}_+^2+\bar{\Lambda}_-^2
+4\Lambda_D^2/3\right)(3l-r_2)}\right]^{1/2}\left(1+\frac{\Delta
r_1}{4l}\right)\left(1+\frac{\Delta r_1}{3l-r_2}\right)^{-1/2}
\\ \label{JB} &&\times F_D^{(2)}\left(1/2;-1,1/2;5/2;\frac{1}{1+\frac{4l}{\Delta
r_1}},\frac{1}{1+\frac{3l-r_2}{\Delta r_1}}\right),\ea \ba\nl J_D
&=& 8T\left[\frac{2l^5\Delta
r_1}{\left(\bar{\Lambda}_+^2+\bar{\Lambda}_-^2
+4\Lambda_D^2/3\right)(3l-r_2)}\right]^{1/2} \left(1+\frac{\Delta
r_1}{3l}\right)\left(1+\frac{\Delta r_1}{3l-r_2}\right)^{-1/2}
\\ \label{JD} &&\times F_D^{(2)}\left(1/2;-2,1/2;3/2;\frac{1}{1+\frac{3l}{\Delta
r_1}},\frac{1}{1+\frac{3l-r_2}{\Delta r_1}}\right).\ea

In the semiclassical limit $r_1\to\infty$, one gets the following
dependence of the energy on the charges $P_I$, $P_{\theta}$ and
$P_{\phi}$ \ba\label{EG3'} E^2=\mathbf{P}^2+2\pi T
\left(6r_0\right)^{1/2}\left(P_{\theta}^2+
\frac{3P_{\phi}^2}{3-\cos^2\theta_2^0}\right)^{1/2}.\ea Again,
this is a generalization of the energy-charge relation $E\sim
K^{1/2}$, and for $\theta_2^0=\pi/2$ (\ref{EG3'}) has the same
form as the expression in (\ref{EG1}).

Now, we turn to the case of string rotating in the background
given in (\ref{2G'}). To satisfy the conditions
$\mathcal{P}_\mu=0$ and the second constraint in (\ref{ecs}), we
use the ansatz \ba\label{se3} X^0=\Lambda_0^0\xi^0, \h
X^I=\Lambda_0^I\xi^0, \h r=r(\xi^1),\h
\phi_1=\Lambda_0^{\phi_1}\xi^0,\h
\phi_2=\Lambda_0^{\phi_2}\xi^0.\ea This embedding is analogous to
the one just considered, where $\theta_1$ is replaced by $\phi_1$.

The first constraint in (\ref{ecs}) takes the form,
\ba\label{ecG2'} &&g_{rr}r'^2 -
\mathcal{U}=\frac{r_0^{1/2}}{C}r'^2-
r_0^{1/2}C\left(v_0^2-\tilde{\Lambda}_+^2A^2-\tilde{\Lambda}_-^2B^2-\tilde{\Lambda}_D^2D^2\right)=0,
\ea where $v_0^2$ is the same as before, and \ba\nl
&&\tilde{\Lambda}_{\pm}^2=\left(\Lambda_0^{\phi_1}\right)^2\sin^2\theta_1^0
+ \left(\Lambda_0^{\phi_2}\right)^2\sin^2\theta_2^0 \pm
2\Lambda_0^{\phi_1}\Lambda_0^{\phi_2}\cos\psi_1^0\sin\theta_1^0\sin\theta_2^0,
\\ \label{LTilde} &&\tilde{\Lambda}_D^2=\left(\Lambda_0^{\phi_1}\cos\theta_1^0
+ \Lambda_0^{\phi_2}\cos\theta_2^0\right)^2.\ea Since
(\ref{ecG2'}) can be obtained from (\ref{ecG3'}) by the
replacements $\bar{\Lambda}_{\pm}^2\to\tilde{\Lambda}_{\mp}^2$,
$\Lambda_D^2\to\tilde{\Lambda}_D^2$, in the same way one can
receive the new values for $r_{max}=r_1$ and $r_2$, the new string
solution from (\ref{ssG3'}), the new expressions for the energy
$E$ and the momenta $P_I$ from (\ref{EP2}). In accordance with
(\ref{scm}), the conserved angular momenta $P_{\phi_1}$ and
$P_{\phi_2}$ are given by \ba\label{Pf1} P_{\phi_1}&=&
\left(\Lambda_0^{\phi_1}\sin\theta_1^0+\Lambda_0^{\phi_2}\cos\psi_1^0\sin\theta_2^0\right)\sin\theta_1^0K_A
\\ \nl
&+&\left(\Lambda_0^{\phi_1}\sin\theta_1^0-\Lambda_0^{\phi_2}\cos\psi_1^0\sin\theta_2^0\right)\sin\theta_1^0K_B
\\ \nl &+&\left(\Lambda_0^{\phi_1}\cos\theta_1^0 +
\Lambda_0^{\phi_2}\cos\theta_2^0\right)\cos\theta_1^0K_D,\ea
\ba\label{Pf2} P_{\phi_2}&=&
\left(\Lambda_0^{\phi_2}\sin\theta_2^0+\Lambda_0^{\phi_1}\cos\psi_1^0\sin\theta_1^0\right)\sin\theta_2^0K_A
\\ \nl
&+&\left(\Lambda_0^{\phi_2}\sin\theta_2^0-\Lambda_0^{\phi_1}\cos\psi_1^0\sin\theta_1^0\right)\sin\theta_2^0K_B
\\ \nl &+&\left(\Lambda_0^{\phi_1}\cos\theta_1^0 +
\Lambda_0^{\phi_2}\cos\theta_2^0\right)\cos\theta_2^0K_D,\ea where
$K_A$, $K_B$, $K_D$ can be obtained from (\ref{JA}), (\ref{JB}),
(\ref{JD}), by the above mentioned replacements.

Taking the semiclassical limit $(r_1\to\infty)$ in the expressions
for $E$, $P_I$, $P_{\phi_1}$ and $P_{\phi_2}$, after some
calculations, one receives the following relation between them
\ba\label{EG2'} E^2&=&\mathbf{P}^2+2\pi T
\left(\frac{6r_0}{\Delta}\right)^{1/2}\times\\ \nl
&&\left[\left(3-\cos^2\theta_2^0\right)P_{\phi_1}^2+
\left(3-\cos^2\theta_1^0\right)P_{\phi_2}^2
-4P_{\phi_1}P_{\phi_2}\cos\theta_1^0\cos\theta_2^0\right]^{1/2},\ea
where \ba\nl \Delta= 3-\cos^2\theta_1^0-
\cos^2\theta_2^0-\cos^2\theta_1^0\cos^2\theta_2^0.\ea This is
another generalization of the energy-charge relation $E\sim
K^{1/2}$, and for $\theta_1^0=\theta_2^0=\pi/2$ has the same form
as the relation in (\ref{EG1}).

The equality (\ref{EG2'}) is only valid for $\Delta\ne 0$. To see
what will be the semiclassical behavior of the rotating string
configuration for $\Delta=0$, let us consider the particular case
$\theta_1^0=\theta_2^0=0$. According to (\ref{Pf1}), (\ref{Pf2}),
the two angular momenta become equal, $P_{\phi_1}=P_{\phi_2}\equiv
P_{\phi}$. Performing the necessary computations, one arrives at
\ba\label{pc2G'} E^2&=&\mathbf{P}^2+6\pi T r_0^{1/2}P_{\phi},\ea
which describes the same type of semiclassical behavior.

Comparing (\ref{se1}), (\ref{se2}) and (\ref{se3}) with each
other, one sees that none of them represents string configuration
with nontrivial wrapping. Then a natural question is if such
solutions do exist at all. The analysis shows that the reason for
the absence of wrapping is that we have too many restrictions on
the embedding parameters $\Lambda_m^\mu$ for the backgrounds
(\ref{1G}), (\ref{3G'}) and (\ref{2G'}). However, it turns out
that if we restrict ourselves to particular cases of these
backgrounds by fixing the values of part of the angles
$\theta_{1,2}^0$, $\psi_1^0$ or $\phi_{1,2}^0$, we can obtain
wrapped rotating string solutions. An example of such solution is
given by the ansatz \ba\nl &&X^0=\Lambda_0^0\xi^0, \h
X^I=\Lambda_0^I\xi^0, \h r=r(\xi^1),\h \theta_1^0=\theta_2^0=0,\\
\nl
&&\psi_1=\Lambda_0^{\psi_1}\xi^0-(\Lambda_1^{\phi_1}+\Lambda_1^{\phi_2})\xi^1,\h
\phi_1=\Lambda_0^{\phi_1}\xi^0+\Lambda_1^{\phi_1}\xi^1,\h
\phi_2=\Lambda_0^{\phi_2}\xi^0+\Lambda_1^{\phi_2}\xi^1.\ea The
background metric felt by the string is \ba\nl ds^2 &=&
r_0^{1/2}\left\{C \left[-(dx^0)^2 + \delta_{IJ}dx^I dx^J +
D^2d(\psi_1+\phi_1+\phi_2)^2\right]+ \frac{dr^2}{C}\right\}.\ea It
can be seen as particular case of (\ref{2G'}) after the
replacement $(\psi_1+\phi_1+\phi_2)\to (\phi_1+\phi_2)$. The
calculations lead to the same result about the semiclassical
behavior of this wrapped string configuration as in (\ref{pc2G'}),
where $P_{\phi}$ must be replaced with $P_{\psi_1}$, $P_{\phi_1}$,
or $P_{\phi_2}$, which are equal to each other.

Another example of wrapped string solution is \ba\nl
&&X^0=\Lambda_0^0\xi^0, \h
X^I=\Lambda_0^I\xi^0, \h r=r(\xi^1),\h \phi_1^0=\theta_2^0=0,\\
\nl &&\theta_1=\Lambda_0^{\theta_1}\xi^0,\h
\psi_1=\Lambda_0^{\psi_1}\xi^0+\Lambda_1^{\psi_1}\xi^1,\h
\phi_2=\Lambda_0^{\phi_2}\xi^0-\Lambda_1^{\psi_1}\xi^1.\ea The
background seen by the string now is \ba\nl ds^2 &=&
r_0^{1/2}\left\{C \left[-(dx^0)^2 + \delta_{IJ}dx^I dx^J
+(A^2+B^2)d\theta_1^2 + D^2d(\psi_1+\phi_2)^2\right]+
\frac{dr^2}{C}\right\},\ea which can be considered as particular
case of (\ref{3G'}) after the replacement $(\psi_1+\phi_2)\to
\phi_2$. In the semiclassical limit, for the above string
configuration, one receives the following energy-charge relation
($P_{\psi_1}=P_{\phi_2}$) \ba\label{pc3G'} &&E^2=\mathbf{P}^2+2\pi
T\left(3r_0\right)^{1/2}\left(2P_{\theta_1}^2+3P_{\psi_1}^2\right)^{1/2}.\ea
It is particular case of (\ref{EG3'}).

Let us finally note that in considering the semiclassical limit
(large charges), we take into account only the leading terms in
the expressions for the conserved quantities. However, there is no
problem to include the higher order terms. For instance, the
inclusion of the next-to-leading order term, modifies
(\ref{pc3G'}) to \ba\label{pc3G'h} &&E^2=\mathbf{P}^2+2\pi
T\left(3r_0\right)^{1/2}\left(2P_{\theta_1}^2+3P_{\psi_1}^2\right)^{1/2}
-\frac{1}{2}(\pi
T)^2(3r_0)^3\frac{P_{\theta_1}^2}{2P_{\theta_1}^2+3P_{\psi_1}^2}.\ea

\setcounter{equation}{0}
\section{Rotating D2-brane solutions, conserved charges
\\and their semiclassical limits}
In this section, we will consider D2-branes rotating in the
backgrounds (\ref{1G}), (\ref{3G'}) and (\ref{2G'}), as it was
already done for strings. It turns out that for every one of these
three backgrounds, there exist two D2-brane configurations of the
type (\ref{sLA}), which ensure that the equalities (\ref{01}),
(\ref{02}) and $\mathcal{P}^2_\mu=0$ are identically satisfied.

We begin with the following D2-brane embedding in the target space
metric (\ref{1G}): \ba\label{DA1} &&X^0=
\Lambda_0^0\xi^0+\frac{\left(\mathbf{\Lambda}_0.\mathbf{\Lambda}_1\right)}{\Lambda_0^0}\left(\xi^1
+ c\xi^2\right), \h X^I=\Lambda_0^I\xi^0 + \Lambda_1^I\left(\xi^1
+ c\xi^2\right),
\\ \nl &&r=r(\xi^2),\h \theta_1=\Lambda_0^{\theta_1}\xi^0,\h \theta_2=\Lambda_0^{\theta_2}\xi^0;
\h
\left(\mathbf{\Lambda}_0.\mathbf{\Lambda}_1\right)=\delta_{IJ}\Lambda_0^I\Lambda_1^J,
\h c=constant.\ea It corresponds to D2-brane extended in the
radial direction $r$, and rotating in the planes given by the
angles $\theta_1$ and $\theta_2$ with constant angular momenta
$P_{\theta_1}$ and $P_{\theta_2}$. It is nontrivially spanned
along $x^0$ and $x^I$ and moves with constant energy $E$, and
constant momenta $P_{I}$.

The metric induced on the D2-brane worldvolume is \ba\nl
&&G_{00}=-r_0^{1/2}C\left(v_0^2 - \Lambda_-^2 A^2-\Lambda_+^2
B^2\right),\\ \nl &&G_{11}=r_0^{1/2}MC,\h G_{12}=cG_{11},\h
G_{22}= g_{rr}r'^2+c^2G_{11},\ea where $v_0^2$ and $\Lambda_{\pm}$
are defined in (\ref{vL1}) and \ba \label{DM} M=
\mathbf{\Lambda}_1^2
-\frac{\left(\mathbf{\Lambda}_0.\mathbf{\Lambda}_1\right)^2}{\left(\Lambda_0^0\right)^2}.\ea
The Lagrangian (\ref{old}) takes the form \ba\nl
&&\mathcal{L}^{A}(\xi^2)
=\frac{1}{4\lambda^0}\left(\tilde{K}_{rr}r'^2 -
\tilde{V}\right),\h \tilde{K}_{rr}=-(2\lambda^0 T_{D2})^2r_0 M e^{-\Phi},\\
\nl &&\tilde{V}=r_{0}^{1/2}C\left(v_0^2 - \Lambda_-^2
A^2-\Lambda_+^2 B^2\right)e^{-\Phi}.\ea From the yet unsolved
constraint (\ref{00e}) \ba\nl \tilde{K}_{rr}r'^2 + \tilde{V}=0,\ea
one obtains the turning points of the effective one-dimensional
periodic motion by solving the equation $r'=0$. In the case under
consideration, the result is given in (\ref{mm}).

Applying the general formula (\ref{D1dc}), we obtain the following
expression for the D2-brane solution \ba\nl
&&\xi^2(r)=\int_{3l}^{r}\left[-\frac{\tilde{K}_{rr}(t)}{\tilde{V}(t)}\right]^{1/2}dt=
\frac{16}{3}\lambda^0 T_{D2}\left[\frac{M l}
{\left(\Lambda_+^2+\Lambda_-^2\right)\left(3l-r_2\right)\Delta
r_1}\right]^{1/2}(2\Delta r)^{3/4}\times
\\ \label{D2s1} &&F_D^{(5)}\left(3/4;-1/4,-1/4,1/4,1/2,1/2;7/4;
-\frac{\Delta r}{2l},-\frac{\Delta r}{4l},-\frac{\Delta r}{6l},
-\frac{\Delta r}{3l-r_2},\frac{\Delta r}{\Delta r_1}\right).\ea

Now, we compute the conserved momenta on the obtained solution
according to (\ref{Dcm}): \ba\label{DEP1}
&&\frac{E}{\Lambda_0^0}=\frac{P_I}{\Lambda_0^I}=
8\pi^2T_{D2}\left[\frac{Ml}{\left(\Lambda_+^2+
\Lambda_-^2\right)(3l-r_2)}\right]^{1/2} \times
\\ \nl &&\left(1+\frac{\Delta
r_1}{2l}\right)^{1/2}\left(1+\frac{\Delta
r_1}{4l}\right)^{1/2}\left(1+\frac{\Delta
r_1}{6l}\right)^{-1/2}\left(1+\frac{\Delta
r_1}{3l-r_2}\right)^{-1/2}\times
\\ \nl && F_D^{(4)}\left(1/2;-1/2,-1/2,1/2,1/2;1;\frac{1}{1+\frac{2l}{\Delta
r_1}},\frac{1}{1+\frac{4l}{\Delta
r_1}},\frac{1}{1+\frac{6l}{\Delta
r_1}},\frac{1}{1+\frac{3l-r_2}{\Delta r_1}}\right),\ea
\ba\label{Dpt12}
&&P_{\theta_1}=\left(\Lambda_0^{\theta_1}-\Lambda_0^{\theta_2}\cos\psi_1^0\right)I_{A1}^D
+\left(\Lambda_0^{\theta_1}+\Lambda_0^{\theta_2}\cos\psi_1^0\right)I_{B1}^D,\\
\nl
&&P_{\theta_2}=\left(\Lambda_0^{\theta_2}-\Lambda_0^{\theta_1}\cos\psi_1^0\right)I_{A1}^D
+\left(\Lambda_0^{\theta_2}+\Lambda_0^{\theta_1}\cos\psi_1^0\right)I_{B1}^D,\ea
where \ba\label{IDA1} I_{A1}^D&=&
8\pi^2T_{D2}\left[\frac{Ml^5}{\left(\Lambda_+^2+
\Lambda_-^2\right)(3l-r_2)}\right]^{1/2} \times
\\ \nl &&\left(1+\frac{\Delta
r_1}{2l}\right)^{3/2}\left(1+\frac{\Delta
r_1}{4l}\right)^{1/2}\left(1+\frac{\Delta
r_1}{6l}\right)^{1/2}\left(1+\frac{\Delta
r_1}{3l-r_2}\right)^{-1/2}\times
\\ \nl && F_D^{(4)}\left(1/2;-3/2,-1/2,-1/2,1/2;1;\frac{1}{1+\frac{2l}{\Delta
r_1}},\frac{1}{1+\frac{4l}{\Delta
r_1}},\frac{1}{1+\frac{6l}{\Delta
r_1}},\frac{1}{1+\frac{3l-r_2}{\Delta r_1}}\right),\ea
\ba\label{IDB1} I_{B1}^D&=&
\frac{4}{3}\pi^2T_{D2}\left[\frac{Ml^3}{\left(\Lambda_+^2+
\Lambda_-^2\right)(3l-r_2)}\right]^{1/2} \times
\\ \nl &&\Delta r_1\left(1+\frac{\Delta
r_1}{2l}\right)^{1/2}\left(1+\frac{\Delta
r_1}{4l}\right)^{3/2}\left(1+\frac{\Delta
r_1}{6l}\right)^{-1/2}\left(1+\frac{\Delta
r_1}{3l-r_2}\right)^{-1/2}\times
\\ \nl && F_D^{(4)}\left(1/2;-1/2,-3/2,1/2,1/2;2;\frac{1}{1+\frac{2l}{\Delta
r_1}},\frac{1}{1+\frac{4l}{\Delta
r_1}},\frac{1}{1+\frac{6l}{\Delta
r_1}},\frac{1}{1+\frac{3l-r_2}{\Delta r_1}}\right).\ea

In the semiclassical limit, (\ref{DEP1}) - (\ref{IDB1}) simplify
to \ba\nl &&\frac{E}{\Lambda_0^0}=\frac{P_I}{\Lambda_0^I}=
\frac{2}{3}\pi^2T_{D2}\left(\frac{M}{\Lambda_+^2+\Lambda_-^2}\right)^{1/2},
\\ \nl  &&P_{\theta_1}=2\Lambda_0^{\theta_1}I_{A1}^D,\h
P_{\theta_2}=2\Lambda_0^{\theta_2}I_{A1}^D,\h
I_{A1}^D=I_{B1}^D=\frac{\sqrt{3}\pi^2T_{D2}M^{1/2}
}{\left(\Lambda_+^2+\Lambda_-^2\right)^{3/2}}v_0^2.\ea From here,
one obtains the following relation between the energy and the
conserved charges \ba\label{ED1}
E^2\left(E^2-\mathbf{P}^2\right)^2 -\frac{2^3}{3^5}(\pi^2
T_{D2})^2\left[\mathbf{\Lambda}_1^2 E^2 -
\left(\mathbf{\Lambda}_1.\mathbf{P}\right)^2\right]
\left(P^2_{\theta_1}+P^2_{\theta_2}\right) = 0,\ea which is third
order algebraic equation for $E^2$. Therefore, this D2-brane
configuration reproduces particular case of the M2-brane
semiclassical behavior given in (4.19) of \cite{B0306} \ba\nl
&&\left\{E^2\left(E^2-\mathbf{P}^2\right) - (2\pi^2 T_{M2}
l_{11}^3)^2\left\{\left(\mathbf{\Lambda}_1\times\mathbf{\Lambda}_2\right)^2
E^2 -
\left[\left(\mathbf{\Lambda}_1\times\mathbf{\Lambda}_2\right)\times\mathbf{P}\right]^2\right\}\right\}^2
\\ \nl &&-6(2\pi^2 T_{M2} l_{11}^3)^2E^2\left[\mathbf{\Lambda}_1^2 E^2 -
\left(\mathbf{\Lambda}_1.\mathbf{P}\right)^2\right]\left(P^2_{\theta}+P^2_{\tilde{\theta}}\right)
= 0,\ea corresponding to
$\left(\mathbf{\Lambda}_1\times\mathbf{\Lambda}_2\right)=0$. For
$\left(\mathbf{\Lambda}_1.\mathbf{P}\right)=0$, (\ref{ED1})
reduces to \ba\nl E^2=\mathbf{P}^2+\frac{2^{3/2}}{3^{5/2}}\pi^2
T_{D2}\mid\mathbf{\Lambda}_1\mid
\left(P^2_{\theta_1}+P^2_{\theta_2}\right)^{1/2}.\ea This is the
same type energy-charge relation as the one obtained for the
string in (\ref{EG1}).

Let us now consider the other possible D2-brane embedding for the
same background metric (\ref{1G}). It is given by \ba\label{DA2}
&&X^0= \Lambda_0^0\xi^0, \h X^I=\Lambda_0^I\xi^0, \h r=r(\xi^2),\\
\nl
&&\theta_1=\Lambda_0^{\theta}\xi^0+\Lambda_1^{\theta}\xi^1+\Lambda_2^{\theta}\xi^2
,\h
\theta_2=\Lambda_0^{\theta}\xi^0-\Lambda_1^{\theta}\xi^1-\Lambda_2^{\theta}\xi^2
.\ea This ansatz describes D2-brane, which is extended along the
radial direction $r$ and rotates in the planes defined by the
angles $\theta_1$ and $\theta_2$, with equal angular momenta
$P_{\theta_1}=P_{\theta_2}=P_{\theta}$. Now we have nontrivial
wrapping along $\theta_1$ and $\theta_2$. In addition, the
D2-brane moves along $x^0$ and $x^I$ with constant energy $E$ and
constant momenta $P_{I}$ respectively.

For the present case, the Lagrangian (\ref{old}) reduces to \ba\nl
&&\mathcal{L}^{A}(\xi^2)
=\frac{1}{4\lambda^0}\left(\tilde{K}_{rr}r'^2 -
\tilde{V}\right),\h \tilde{K}_{rr}=-(2\lambda^0 T_{D2})^2r_0
\left(\Lambda_{1+}^2A^2
+ \Lambda_{1-}^2B^2\right) e^{-\Phi},\\
\nl &&\tilde{V}=r_{0}^{1/2}C\left(v_0^2 - \check{\Lambda}_-^2
A^2-\check{\Lambda}_+^2 B^2\right)e^{-\Phi},\ea where \ba\nl
\Lambda_{1\pm}^2=2\left(\Lambda_1^{\theta}\right)^2\left(1\pm\cos\psi_1^0\right),\h
\check{\Lambda}_{\pm}^2=2\left(\Lambda_0^{\theta}\right)^2\left(1\pm\cos\psi_1^0\right).\ea
The constraint (\ref{00e}) \ba\nl \tilde{K}_{rr}r'^2 +
\tilde{V}=0,\ea leads to the same solutions of the equation
$r'=0$, as given in (\ref{mm}), but in terms of the new parameters
$\check{\Lambda}_{\pm}$ instead of $\Lambda_{\pm}$.

Replacing the above expressions for $\tilde{K}_{rr}$ and
$\tilde{V}$ in (\ref{D1dc}), we obtain the D2-brane solution:
\ba\nl &&\xi^2(r)=\frac{8}{3}\lambda^0
T_{D2}\left[\frac{l\left(\Lambda_{1+}^2+\Lambda_{1-}^2\right)(3l-v_+)(3l-v_-)}
{3\left(\check{\Lambda}_+^2+\check{\Lambda}_-^2\right)\left(3l-r_2\right)\Delta
r_1}\right]^{1/2}(2\Delta r)^{3/4}\times
\\ \label{D2s2} &&F_D^{(7)}\left(3/4;-1/4,-1/4,1/4,-1/2,-1/2,1/2,1/2;7/4;\right.\\
\nl &&\left.-\frac{\Delta r}{2l},-\frac{\Delta
r}{4l},-\frac{\Delta r}{6l}, -\frac{\Delta
r}{3l-v_+},-\frac{\Delta r}{3l-v_-}, -\frac{\Delta
r}{3l-r_2},\frac{\Delta r}{\Delta r_1}\right),\ea where $v_{\pm}$
are the zeros of the polynomial \ba\nl
t^2-2l\frac{\Lambda_{1+}^2-\Lambda_{1-}^2}{\Lambda_{1+}^2+\Lambda_{1-}^2}t-3l^2=(t-v_+)(t-v_-).\ea

In the case under consideration, the conserved quantities are $E$,
${P_I}$ and $P_{\theta}$. By using (\ref{Dcm}), we derive the
following result for them \ba\label{DEP2}
&&\frac{E}{\Lambda_0^0}=\frac{P_I}{\Lambda_0^I}=
4\pi^2T_{D2}\left[\frac{l\left(\Lambda_{1+}^2+\Lambda_{1-}^2\right)
(3l-v_+)(3l-v_-)}{3\left(\check{\Lambda}_{+}^2+\check{\Lambda}_{-}^2\right)
(3l-r_2)}\right]^{1/2}\times
\\ \nl &&\left(1+\frac{\Delta
r_1}{2l}\right)^{1/2}\left(1+\frac{\Delta
r_1}{4l}\right)^{1/2}\left(1+\frac{\Delta r_1}{6l}\right)^{-1/2}\times \\
\nl &&\left(1+\frac{\Delta r_1}{3l-v_+}\right)^{1/2}
\left(1+\frac{\Delta r_1}{3l-v_-}\right)^{1/2}\left(1+\frac{\Delta
r_1}{3l-r_2}\right)^{-1/2}\times
\\ \nl && F_D^{(6)}\left(1/2;-1/2,-1/2,1/2,-1/2,-1/2,1/2;1;\right.
\\ \nl &&\left.\frac{1}{1+\frac{2l}{\Delta
r_1}},\frac{1}{1+\frac{4l}{\Delta
r_1}},\frac{1}{1+\frac{6l}{\Delta
r_1}},\frac{1}{1+\frac{3l-v+}{\Delta
r_1}},\frac{1}{1+\frac{3l-v-}{\Delta
r_1}}\frac{1}{1+\frac{3l-r_2}{\Delta r_1}}\right),\ea \ba\nl
P_{\theta}=
\Lambda_0^{\theta}\left[\left(1-\cos\psi_1^0\right)I_{A2}^D +
\left(1+\cos\psi_1^0\right)I_{B2}^D\right],\ea where \ba\nl
&&I_{A2}^D=
4\pi^2T_{D2}\left[\frac{l^5\left(\Lambda_{1+}^2+\Lambda_{1-}^2\right)
(3l-v_+)(3l-v_-)}{3\left(\check{\Lambda}_{+}^2+\check{\Lambda}_{-}^2\right)
(3l-r_2)}\right]^{1/2}\times
\\ \nl &&\left(1+\frac{\Delta
r_1}{2l}\right)^{3/2}\left(1+\frac{\Delta
r_1}{4l}\right)^{1/2}\left(1+\frac{\Delta r_1}{6l}\right)^{1/2}\times \\
\nl &&\left(1+\frac{\Delta r_1}{3l-v_+}\right)^{1/2}
\left(1+\frac{\Delta r_1}{3l-v_-}\right)^{1/2}\left(1+\frac{\Delta
r_1}{3l-r_2}\right)^{-1/2}\times
\\ \nl && F_D^{(6)}\left(1/2;-3/2,-1/2,-1/2,-1/2,-1/2,1/2;1;\right.
\\ \nl &&\left.\frac{1}{1+\frac{2l}{\Delta
r_1}},\frac{1}{1+\frac{4l}{\Delta
r_1}},\frac{1}{1+\frac{6l}{\Delta
r_1}},\frac{1}{1+\frac{3l-v+}{\Delta
r_1}},\frac{1}{1+\frac{3l-v-}{\Delta
r_1}}\frac{1}{1+\frac{3l-r_2}{\Delta r_1}}\right),\ea \ba\nl
&&I_{B2}^D=
2\pi^2T_{D2}\left[\frac{l^3\left(\Lambda_{1+}^2+\Lambda_{1-}^2\right)
(3l-v_+)(3l-v_-)}{3^3\left(\check{\Lambda}_{+}^2+\check{\Lambda}_{-}^2\right)
(3l-r_2)}\right]^{1/2}\times
\\ \nl &&\Delta r_1\left(1+\frac{\Delta
r_1}{2l}\right)^{1/2}\left(1+\frac{\Delta
r_1}{4l}\right)^{3/2}\left(1+\frac{\Delta r_1}{6l}\right)^{-1/2}\times \\
\nl &&\left(1+\frac{\Delta r_1}{3l-v_+}\right)^{1/2}
\left(1+\frac{\Delta r_1}{3l-v_-}\right)^{1/2}\left(1+\frac{\Delta
r_1}{3l-r_2}\right)^{-1/2}\times
\\ \nl && F_D^{(6)}\left(1/2;-1/2,-3/2,1/2,-1/2,-1/2,1/2;2;\right.
\\ \nl &&\left.\frac{1}{1+\frac{2l}{\Delta
r_1}},\frac{1}{1+\frac{4l}{\Delta
r_1}},\frac{1}{1+\frac{6l}{\Delta
r_1}},\frac{1}{1+\frac{3l-v+}{\Delta
r_1}},\frac{1}{1+\frac{3l-v-}{\Delta
r_1}}\frac{1}{1+\frac{3l-r_2}{\Delta r_1}}\right).\ea

Taking the semiclassical limit in the above
expressions\footnote{In this limit $v_{\pm}$ remain finite.}, we
obtain the following dependence of the energy on $P_I$ and
$P_{\theta}$: \ba\label{ED2} E^2=\mathbf{P}^2+3^{5/3}(2\pi T_{D2}
\Lambda_1^{\theta})^{2/3}P_{\theta}^{4/3}.\ea This is the same
semiclassical behavior as the one exhibited by the M2-brane as
given in (4.27) of \cite{B0306}: \ba\nl
E^2=\mathbf{P}^2+3^{5/3}(2\pi T_{M2}
l_{11}^3\Lambda_1^6)^{2/3}P_{\theta}^{4/3},\ea which is a
generalization of the energy-charge relation $E\sim K^{2/3}$ for
the case $\mathbf{P}\ne 0$.

Now, we turn to the case of D2-branes rotating in the background
(\ref{3G'}). Again, we have two possible embeddings of the type
(\ref{sLA}). The first one is given by the ansatz \ba\label{DA3}
&&X^0=
\Lambda_0^0\xi^0+\frac{\left(\mathbf{\Lambda}_0.\mathbf{\Lambda}_1\right)}{\Lambda_0^0}\left(\xi^1
+ c\xi^2\right), \h X^I=\Lambda_0^I\xi^0 + \Lambda_1^I\left(\xi^1
+ c\xi^2\right),
\\ \nl &&r=r(\xi^2),\h \theta_1=\Lambda_0^{\theta}\xi^0,\h
\phi_2=\Lambda_0^{\phi}\xi^0.\ea (\ref{DA3}) is analogous to
(\ref{DA1}), but now the rotations are in the planes defined by
the angles $\theta_1$ and $\phi_2$ instead of $\theta_1$ and
$\theta_2$.

The Lagrangian (\ref{old}) takes the form \ba\nl
&&\mathcal{L}^{A}(\xi^2)
=\frac{1}{4\lambda^0}\left(\tilde{K}_{rr}r'^2 -
\tilde{V}\right),\h \tilde{K}_{rr}=-(2\lambda^0 T_{D2})^2r_0 M e^{-\Phi},\\
\nl &&\tilde{V}=r_{0}^{1/2}C\left(v_0^2 - \bar{\Lambda}_-^2
A^2-\bar{\Lambda}_+^2 B^2-\Lambda_D^2 D^2\right)e^{-\Phi},\ea
where $M$, $v_0^2$, $\bar{\Lambda}_\pm^2$ and $\Lambda_D^2$ are
defined in (\ref{DM}), (\ref{vL1}) and (\ref{LBar}) respectively.

The solution $\xi^2(r)$ can be obtained from (\ref{D2s1}) by the
replacement \ba\label{R1}\Lambda_+^2+\Lambda_-^2\to
\bar{\Lambda}_+^2+\bar{\Lambda}_-^2+4\Lambda_D^2/3.\ea It is
understood, that the solutions $r_{max}=r_1$ and $r_2$ of $r'=0$
are also correspondingly changed ($r_{min}$ remains the same). The
explicit expressions for $E$ and $P_I$ can be obtained in the same
way from (\ref{DEP1}). The computation of the conserved angular
momenta $P_{\theta}$ and $P_{\phi}$ according to (\ref{Dcm}) gives
\ba\nl
P_{\theta}&=&\left(\Lambda_0^{\theta}-\Lambda_0^{\phi}\sin\psi_1^0\sin\theta_2^0\right)J_{A1}^D
+\left(\Lambda_0^{\theta}+\Lambda_0^{\phi}\sin\psi_1^0\sin\theta_2^0\right)J_{B1}^D,
\\ \nl
P_{\phi}&=&\left(\Lambda_0^{\phi}\sin\theta_2^0-\Lambda_0^{\theta}\sin\psi_1^0\right)\sin\theta_2^0J_{A1}^D
+\left(\Lambda_0^{\phi}\sin\theta_2^0+\Lambda_0^{\theta}\sin\psi_1^0\right)\sin\theta_2^0J_{B1}^D
\\ \nl &&+\Lambda_0^{\phi}\cos^2\theta_2^0 J_{D1}^D, \ea where one
obtains $J_{A1}^D$, $J_{B1}^D$ from (\ref{IDA1}), (\ref{IDB1}) by
the replacement (\ref{R1}), and \ba\nl J_{D1}^D&=&
8\pi^2T_{D2}\left[\frac{Ml^5}{\left(\bar{\Lambda}_+^2+
\bar{\Lambda}_-^2+4\Lambda_D^2/3\right)(3l-r_2)}\right]^{1/2}
\times
\\ \nl &&\left(1+\frac{\Delta
r_1}{2l}\right)^{1/2}\left(1+\frac{\Delta
r_1}{3l}\right)^{2}\left(1+\frac{\Delta
r_1}{4l}\right)^{1/2}\left(1+\frac{\Delta
r_1}{6l}\right)^{-1/2}\left(1+\frac{\Delta
r_1}{3l-r_2}\right)^{-1/2}\times
\\ \nl && F_D^{(5)}\left(1/2;-1/2,-2,-1/2,1/2,1/2;1;\right.
\\ \nl &&\left.\frac{1}{1+\frac{2l}{\Delta
r_1}},\frac{1}{1+\frac{3l}{\Delta
r_1}},\frac{1}{1+\frac{4l}{\Delta
r_1}},\frac{1}{1+\frac{6l}{\Delta
r_1}},\frac{1}{1+\frac{3l-r_2}{\Delta r_1}}\right).\ea

Taking $r_1\to\infty$ in the above expressions, one obtains that
in the semiclassical limit the following energy-charge relation
holds \ba\nl \frac{E^2\left(
E^2-\mathbf{P}^2\right)^2}{\mathbf{\Lambda}_1^2 E^2
-\left(\mathbf{\Lambda}_1.\mathbf{P}\right)^2}
=\frac{2^3}{3^5}(\pi^2 T_{D2})^2
\left(P^2_{\theta}+\frac{3P_{\phi}^2}{3-\cos^2\theta_2^0}\right).\ea
Obviously, this is a generalization of the relation (\ref{ED1})
and for $\theta_2^0=\pi/2$ has the same form.

Let us see if another D2-brane embedding for the target space
metric (\ref{3G'}) is possible. It turns out that in this case
such nontrivial solution exists if the non-diagonal part of the
metric (\ref{3G'}) is absent. Otherwise, we have too many
conditions on the embedding parameters, which leads to vanishing
kinetic term in the Lagrangian (\ref{old}): $\tilde{K}_{rr}=0$.
That is why, we will consider the particular case $\psi_1^0=0$.
Then, the other possible ansatz is \ba\label{DA4} X^0=
\Lambda_0^0\xi^0, \h X^I=\Lambda_0^I\xi^0, \h r=r(\xi^2),\h
\theta_1=\Lambda_1^{\theta}\xi^1+\Lambda_2^{\theta}\xi^2 ,\h
\phi_2=\Lambda_0^{\phi}\xi^0 ,\ea i.e., we have D2-brane extended
in the radial direction $r$, wrapped along the angular coordinate
$\theta_1$ and rotating in the plane given by the angle $\phi_2$.
The embedding \ba\nl X^0= \Lambda_0^0\xi^0, \h
X^I=\Lambda_0^I\xi^0, \h r=r(\xi^2),\h
\theta_1=\Lambda_0^{\theta}\xi^0 ,\h
\phi_2=\Lambda_1^{\phi}\xi^1+\Lambda_2^{\phi}\xi^2 \ea is also
admissible, but it just interchanges the role of the angles
$\theta_1$ and $\phi_2$.

For the ansatz (\ref{DA4}), the Lagrangian (\ref{old}) is given by
\ba\nl &&\mathcal{L}^{A}(\xi^2)
=\frac{1}{4\lambda^0}\left(\tilde{K}_{rr}r'^2 -
\tilde{V}\right),\h \tilde{K}_{rr}=-(2\lambda^0 T_{D2})^2r_0
(\Lambda_1^{\theta})^2\left(A^2+B^2\right) e^{-\Phi},\\
\nl &&\tilde{V}=r_{0}^{1/2}C\left[v_0^2 -
\Lambda^2\left(A^2+B^2\right) -\Lambda_D^2 D^2\right]e^{-\Phi},\ea
where \ba\nl \Lambda^2=(\Lambda_0^{\phi})^2\sin^2\theta_2^0.\ea
$v_0^2$ and $\Lambda_D^2$ are introduced in (\ref{vL1}) and
(\ref{LBar}) respectively. The solutions of the equation $r'=0$
determining the turning points of the periodic motion now are:
\ba\nl r_{min}=3l,\h
r_{max}=r_1=3\sqrt{\frac{2v_0^2}{3\Lambda^2+2\Lambda_D^2}}=-r_2.\ea

Replacing the above expressions for $\tilde{K}_{rr}$ and
$\tilde{V}$ in (\ref{D1dc}), one obtains the solution: \ba\nl
&&\xi^2(r)=\frac{8}{3}\lambda^0 T_{D2}\Lambda_1^{\theta}
\left[\frac{l(3l-w_+)(3l-w_-)}
{\left(3\Lambda^2+2\Lambda_D^2\right)\left(3l-r_2\right)\Delta
r_1}\right]^{1/2}(2\Delta r)^{3/4}\times
\\ \label{D2s4} &&F_D^{(7)}\left(3/4;-1/4,-1/4,1/4,-1/2,-1/2,1/2,1/2;7/4;\right.\\
\nl &&\left.-\frac{\Delta r}{2l},-\frac{\Delta
r}{4l},-\frac{\Delta r}{6l}, -\frac{\Delta
r}{3l-w_+},-\frac{\Delta r}{3l-w_-}, -\frac{\Delta
r}{3l-r_2},\frac{\Delta r}{\Delta r_1}\right),\h
w_{\pm}=\pm\sqrt{3} l.\ea The computation of the conserved
quantities $E$, $P_I$ and $P_{\phi_2}\equiv P_{\phi}$, in
accordance with (\ref{Dcm}), gives \ba\label{DEP4}
&&\frac{E}{\Lambda_0^0}=\frac{P_I}{\Lambda_0^I}= 4\pi^2
T_{D2}\Lambda_1^{\theta} \left[\frac{l(3l-w_+)(3l-w_-)}
{\left(3\Lambda^2+2\Lambda_D^2\right)\left(3l-r_2\right)}\right]^{1/2}
\times\\ \nl &&\left(1+\frac{\Delta
r_1}{2l}\right)^{1/2}\left(1+\frac{\Delta
r_1}{4l}\right)^{1/2}\left(1+\frac{\Delta r_1}{6l}\right)^{-1/2}\times \\
\nl &&\left(1+\frac{\Delta r_1}{3l-w_+}\right)^{1/2}
\left(1+\frac{\Delta r_1}{3l-w_-}\right)^{1/2}\left(1+\frac{\Delta
r_1}{3l-r_2}\right)^{-1/2}\times
\\ \nl &&F_D^{(6)}\left(1/2;-1/2,-1/2,1/2,-1/2,-1/2,1/2;1;\right.\\
\nl &&\left.\frac{1}{1+\frac{2l}{\Delta
r_1}},\frac{1}{1+\frac{4l}{\Delta
r_1}},\frac{1}{1+\frac{6l}{\Delta
r_1}},\frac{1}{1+\frac{3l-w+}{\Delta
r_1}},\frac{1}{1+\frac{3l-w-}{\Delta
r_1}}\frac{1}{1+\frac{3l-r_2}{\Delta r_1}}\right),\ea \ba\nl
P_\phi=\sin^2\theta_2^0\left(J_{A2}^D+J_{B2}^D\right)
+\cos^2\theta_2^0 J_{D2}^D,\ea where \ba\nl && J_{A2}^D=4\pi^2
T_{D2}\Lambda_0^\phi \Lambda_1^{\theta}
\left[\frac{l^5(3l-w_+)(3l-w_-)}
{\left(3\Lambda^2+2\Lambda_D^2\right)\left(3l-r_2\right)}\right]^{1/2}
\times\\ \nl &&\left(1+\frac{\Delta
r_1}{2l}\right)^{3/2}\left(1+\frac{\Delta
r_1}{4l}\right)^{1/2}\left(1+\frac{\Delta r_1}{6l}\right)^{1/2}\times \\
\nl &&\left(1+\frac{\Delta r_1}{3l-w_+}\right)^{1/2}
\left(1+\frac{\Delta r_1}{3l-w_-}\right)^{1/2}\left(1+\frac{\Delta
r_1}{3l-r_2}\right)^{-1/2}\times
\\ \nl &&F_D^{(6)}\left(1/2;-3/2,-1/2,-1/2,-1/2,-1/2,1/2;1;\right.\\
\nl &&\left.\frac{1}{1+\frac{2l}{\Delta
r_1}},\frac{1}{1+\frac{4l}{\Delta
r_1}},\frac{1}{1+\frac{6l}{\Delta
r_1}},\frac{1}{1+\frac{3l-w+}{\Delta
r_1}},\frac{1}{1+\frac{3l-w-}{\Delta
r_1}}\frac{1}{1+\frac{3l-r_2}{\Delta r_1}}\right),\ea \ba\nl &&
J_{B2}^D=\frac{2}{3}\pi^2 T_{D2}\Lambda_0^\phi \Lambda_1^{\theta}
\left[\frac{l^3(3l-w_+)(3l-w_-)}
{\left(3\Lambda^2+2\Lambda_D^2\right)\left(3l-r_2\right)}\right]^{1/2}
\times\\ \nl &&\Delta r_1\left(1+\frac{\Delta
r_1}{2l}\right)^{1/2}\left(1+\frac{\Delta
r_1}{4l}\right)^{3/2}\left(1+\frac{\Delta r_1}{6l}\right)^{-1/2}\times \\
\nl &&\left(1+\frac{\Delta r_1}{3l-w_+}\right)^{1/2}
\left(1+\frac{\Delta r_1}{3l-w_-}\right)^{1/2}\left(1+\frac{\Delta
r_1}{3l-r_2}\right)^{-1/2}\times
\\ \nl &&F_D^{(6)}\left(1/2;-1/2,-3/2,1/2,-1/2,-1/2,1/2;2;\right.\\
\nl &&\left.\frac{1}{1+\frac{2l}{\Delta
r_1}},\frac{1}{1+\frac{4l}{\Delta
r_1}},\frac{1}{1+\frac{6l}{\Delta
r_1}},\frac{1}{1+\frac{3l-w+}{\Delta
r_1}},\frac{1}{1+\frac{3l-w-}{\Delta
r_1}}\frac{1}{1+\frac{3l-r_2}{\Delta r_1}}\right),\ea \ba\nl &&
J_{D2}^D=4\pi^2 T_{D2}\Lambda_0^\phi \Lambda_1^{\theta}
\left[\frac{l^5(3l-w_+)(3l-w_-)}
{\left(3\Lambda^2+2\Lambda_D^2\right)\left(3l-r_2\right)}\right]^{1/2}
\times\\ \nl &&\left(1+\frac{\Delta
r_1}{2l}\right)^{1/2}\left(1+\frac{\Delta
r_1}{3l}\right)^{2}\left(1+\frac{\Delta
r_1}{4l}\right)^{1/2}\left(1+\frac{\Delta r_1}{6l}\right)^{-1/2}\times \\
\nl &&\left(1+\frac{\Delta r_1}{3l-w_+}\right)^{1/2}
\left(1+\frac{\Delta r_1}{3l-w_-}\right)^{1/2}\left(1+\frac{\Delta
r_1}{3l-r_2}\right)^{-1/2}\times
\\ \nl &&F_D^{(7)}\left(1/2;-1/2,-2,-1/2,1/2,-1/2,-1/2,1/2;1;\right.\\
\nl &&\left.\frac{1}{1+\frac{2l}{\Delta
r_1}},\frac{1}{1+\frac{3l}{\Delta
r_1}},\frac{1}{1+\frac{4l}{\Delta
r_1}},\frac{1}{1+\frac{6l}{\Delta
r_1}},\frac{1}{1+\frac{3l-w+}{\Delta
r_1}},\frac{1}{1+\frac{3l-w-}{\Delta
r_1}}\frac{1}{1+\frac{3l-r_2}{\Delta r_1}}\right).\ea

Going to the semiclassical limit $r_1\to\infty$ in the above
expressions for the conserved quantities, one obtains the
following relation between them \ba\label{ED4}
E^2=\mathbf{P}^2+\frac{3^{7/3}}{2^{1/3}}\left(\frac{\pi T_{D2}
\Lambda_1^{\theta}}{3-\cos^2\theta_2^0}\right)^{2/3}P_{\phi}^{4/3}.\ea
This is a generalization of the energy-charge relation received in
(\ref{ED2}).

Our next task is to consider D2-branes rotating in the background
(\ref{2G'}). One admissible embedding is \ba\label{DA5} &&X^0=
\Lambda_0^0\xi^0+\frac{\left(\mathbf{\Lambda}_0.\mathbf{\Lambda}_1\right)}{\Lambda_0^0}\left(\xi^1
+ c\xi^2\right), \h X^I=\Lambda_0^I\xi^0 + \Lambda_1^I\left(\xi^1
+ c\xi^2\right),
\\ \nl &&r=r(\xi^2),\h \phi_1=\Lambda_0^{\phi_1}\xi^0,\h
\phi_2=\Lambda_0^{\phi_2}\xi^0.\ea It is analogous to (\ref{DA1})
and (\ref{DA3}), but now the rotations are in the planes given by
the angles $\phi_1$ and $\phi_2$.

The D2-brane Lagrangian (\ref{old}) now reads \ba\nl
&&\mathcal{L}^{A}(\xi^2)
=\frac{1}{4\lambda^0}\left(\tilde{K}_{rr}r'^2 -
\tilde{V}\right),\h \tilde{K}_{rr}=-(2\lambda^0 T_{D2})^2r_0 M e^{-\Phi},\\
\nl &&\tilde{V}=r_{0}^{1/2}C\left(v_0^2 - \tilde{\Lambda}_+^2
A^2-\tilde{\Lambda}_-^2 B^2-\tilde{\Lambda}_D^2
D^2\right)e^{-\Phi},\ea where $M$, $v_0^2$,
$\tilde{\Lambda}_\pm^2$ and $\tilde{\Lambda}_D^2$ are defined in
(\ref{DM}), (\ref{vL1}) and (\ref{LTilde}) respectively. The
values for $r_{max}=r_1$ and $r_2$, the solution $\xi^2(r)$, and
the expressions for $E$, $P_I$, may be obtained from the
corresponding quantities for the embedding (\ref{DA3}) by the
replacements $\bar{\Lambda}_{\mp}^2\to\tilde{\Lambda}_{\pm}^2$,
$\Lambda_D^2\to \tilde{\Lambda}_D^2$. For the conserved angular
momenta $P_{\phi_1}$ and $P_{\phi_2}$, (\ref{Dcm}) gives
\ba\label{D5Pf1} P_{\phi_1}&=&
\left(\Lambda_0^{\phi_1}\sin\theta_1^0+\Lambda_0^{\phi_2}\cos\psi_1^0\sin\theta_2^0\right)\sin\theta_1^0K_{A1}^D
\\ \nl
&+&\left(\Lambda_0^{\phi_1}\sin\theta_1^0-\Lambda_0^{\phi_2}\cos\psi_1^0\sin\theta_2^0\right)\sin\theta_1^0K_{B1}^D
\\ \nl &+&\left(\Lambda_0^{\phi_1}\cos\theta_1^0 +
\Lambda_0^{\phi_2}\cos\theta_2^0\right)\cos\theta_1^0K_{D1}^D,\ea
\ba\label{D5Pf2} P_{\phi_2}&=&
\left(\Lambda_0^{\phi_2}\sin\theta_2^0+\Lambda_0^{\phi_1}\cos\psi_1^0\sin\theta_1^0\right)\sin\theta_2^0K_{A1}^D
\\ \nl
&+&\left(\Lambda_0^{\phi_2}\sin\theta_2^0-\Lambda_0^{\phi_1}\cos\psi_1^0\sin\theta_1^0\right)\sin\theta_2^0K_{B1}^D
\\ \nl &+&\left(\Lambda_0^{\phi_1}\cos\theta_1^0 +
\Lambda_0^{\phi_2}\cos\theta_2^0\right)\cos\theta_2^0K_{D1}^D,\ea
where $K_{A1}^D$, $K_{B1}^D$ and $K_{D1}^D$ can be obtained from
$J_{A1}^D$, $J_{B1}^D$ and $J_{D1}^D$ through the above mentioned
replacements.

The calculations show that in the semiclassical limit, the
dependence of the energy on the conserved charges, for the present
case, is given by the equality: \ba\label{ED5} &&\frac{E^2\left(
E^2-\mathbf{P}^2\right)^2}{\mathbf{\Lambda}_1^2 E^2
-\left(\mathbf{\Lambda}_1.\mathbf{P}\right)^2} =\\ \nl
&&\frac{2^3}{3^5}(\pi^2
T_{D2})^2\frac{\left(3-\cos^2\theta_2^0\right)P_{\phi_1}^2+
\left(3-\cos^2\theta_1^0\right)P_{\phi_2}^2
-4P_{\phi_1}P_{\phi_2}\cos\theta_1^0\cos\theta_2^0}
{3-\cos^2\theta_1^0-
\cos^2\theta_2^0-\cos^2\theta_1^0\cos^2\theta_2^0}.\ea This is
another generalization of the energy-charge relation $E\sim
K^{1/2}$, and for $\theta_1^0=\theta_2^0=\pi/2$ has the same form
as the relation in (\ref{ED1}).

Finally, let us consider the other possible D2-brane embedding in
the background (\ref{2G'}). It turns out that such nontrivial
embedding do exists only for $\theta_1^0=\theta_2^0\equiv
\theta^0$, and is given by the ansatz \ba\label{DA6}
&&X^0= \Lambda_0^0\xi^0, \h X^I=\Lambda_0^I\xi^0, \h r=r(\xi^2),\\
\nl
&&\phi_1=\Lambda_0^{\phi}\xi^0+\Lambda_1^{\phi}\xi^1+\Lambda_2^{\phi}\xi^2
,\h
\phi_2=\Lambda_0^{\phi}\xi^0-\Lambda_1^{\phi}\xi^1-\Lambda_2^{\phi}\xi^2
.\ea It describes D2-brane configuration, which is analogous to
the one in (\ref{DA2}), but now the rotations are in the planes
defined by the angles $\phi_1$ and $\phi_2$ instead of $\theta_1$
and $\theta_2$.

For this embedding, the Lagrangian (\ref{old}) have the form
\ba\nl &&\mathcal{L}^{A}(\xi^2)
=\frac{1}{4\lambda^0}\left(\tilde{K}_{rr}r'^2 -
\tilde{V}\right),\h \tilde{K}_{rr}=-(2\lambda^0 T_{D2})^2r_0
\left(\hat{\Lambda}_{1-}^2A^2+\hat{\Lambda}_{1+}^2B^2\right) e^{-\Phi},\\
\nl &&\tilde{V}=r_{0}^{1/2}C\left(v_0^2 - \hat{\Lambda}_+^2
A^2-\hat{\Lambda}_-^2 B^2-\hat{\Lambda}_D^2
D^2\right)e^{-\Phi},\ea where $v_0^2$ is defined in (\ref{vL1})
and \ba\nl
&&\hat{\Lambda}_{1\pm}^2=2(1\pm\cos\psi_1^0)\sin^2\theta^0(\Lambda_1^\phi)^2,\\
\nl
&&\hat{\Lambda}_\pm^2=2(1\pm\cos\psi_1^0)\sin^2\theta^0(\Lambda_0^\phi)^2,\\
\nl &&\hat{\Lambda}_D^2=4\cos^2\theta^0(\Lambda_0^\phi)^2.\ea The
constraint (\ref{00e}), $\tilde{K}_{rr}r'^2 + \tilde{V}=0$, leads
to the same solutions of the equation $r'=0$, as for the case just
considered, but in terms of the new parameters
$\hat{\Lambda}_{\pm}^2$, $\hat{\Lambda}_{D}^2$.

In accordance with (\ref{D1dc}), one obtains \ba\nl
&&\xi^2(r)=\frac{8}{3}\lambda^0
T_{D2}\left[\frac{l\left(\hat{\Lambda}_{1+}^2+\hat{\Lambda}_{1-}^2\right)(3l-u_+)(3l-u_-)}
{3\left(\hat{\Lambda}_+^2+\hat{\Lambda}_-^2+4\hat{\Lambda}_D^2/3\right)\left(3l-r_2\right)\Delta
r_1}\right]^{1/2}(2\Delta r)^{3/4}\times
\\ \nl &&F_D^{(7)}\left(3/4;-1/4,-1/4,1/4,-1/2,-1/2,1/2,1/2;7/4;\right.\\
\nl &&\left.-\frac{\Delta r}{2l},-\frac{\Delta
r}{4l},-\frac{\Delta r}{6l}, -\frac{\Delta
r}{3l-u_+},-\frac{\Delta r}{3l-u_-}, -\frac{\Delta
r}{3l-r_2},\frac{\Delta r}{\Delta r_1}\right),\ea where \ba\nl
u_\pm=l\left[\frac{\hat{\Lambda}_{1+}^2-\hat{\Lambda}_{1-}^2}
{\hat{\Lambda}_{1+}^2+\hat{\Lambda}_{1-}^2}\pm
\sqrt{3+\left(\frac{\hat{\Lambda}_{1+}^2-\hat{\Lambda}_{1-}^2}
{\hat{\Lambda}_{1+}^2+\hat{\Lambda}_{1-}^2}\right)^2}\right].\ea

The computation of the conserved charges (\ref{Dcm}) results in
\ba\label{DEP6} &&\frac{E}{\Lambda_0^0}=\frac{P_I}{\Lambda_0^I}=
4\pi^2T_{D2}\left[\frac{l\left(\hat{\Lambda}_{1+}^2+\hat{\Lambda}_{1-}^2\right)(3l-u_+)(3l-u_-)}
{3\left(\hat{\Lambda}_+^2+\hat{\Lambda}_-^2+4\hat{\Lambda}_D^2/3\right)\left(3l-r_2\right)}\right]^{1/2}\times
\\ \nl &&\left(1+\frac{\Delta
r_1}{2l}\right)^{1/2}\left(1+\frac{\Delta
r_1}{4l}\right)^{1/2}\left(1+\frac{\Delta r_1}{6l}\right)^{-1/2}\times \\
\nl &&\left(1+\frac{\Delta r_1}{3l-u_+}\right)^{1/2}
\left(1+\frac{\Delta r_1}{3l-u_-}\right)^{1/2}\left(1+\frac{\Delta
r_1}{3l-r_2}\right)^{-1/2}\times
\\ \nl && F_D^{(6)}\left(1/2;-1/2,-1/2,1/2,-1/2,-1/2,1/2;1;\right.
\\ \nl &&\left.\frac{1}{1+\frac{2l}{\Delta
r_1}},\frac{1}{1+\frac{4l}{\Delta
r_1}},\frac{1}{1+\frac{6l}{\Delta
r_1}},\frac{1}{1+\frac{3l-u+}{\Delta
r_1}},\frac{1}{1+\frac{3l-u-}{\Delta
r_1}}\frac{1}{1+\frac{3l-r_2}{\Delta r_1}}\right),\ea \ba\nl
&&P_{\phi} \equiv P_{\phi_1}=P_{\phi_2}=\\
\nl
&&\Lambda_0^\phi\left\{\sin^2\theta^0\left[\left(1+\cos\psi_1^0\right)K^D_{A2}
+\left(1-\cos\psi_1^0\right)K^D_{B2}\right]+2\cos^2\theta^0K^D_{D2}\right\},\ea
where \ba\nl &&K^D_{A2}=
4\pi^2T_{D2}\left[\frac{l^5\left(\hat{\Lambda}_{1+}^2+\hat{\Lambda}_{1-}^2\right)(3l-u_+)(3l-u_-)}
{3\left(\hat{\Lambda}_+^2+\hat{\Lambda}_-^2+4\hat{\Lambda}_D^2/3\right)\left(3l-r_2\right)}\right]^{1/2}\times
\\ \nl &&\left(1+\frac{\Delta
r_1}{2l}\right)^{3/2}\left(1+\frac{\Delta
r_1}{4l}\right)^{1/2}\left(1+\frac{\Delta r_1}{6l}\right)^{1/2}\times \\
\nl &&\left(1+\frac{\Delta r_1}{3l-u_+}\right)^{1/2}
\left(1+\frac{\Delta r_1}{3l-u_-}\right)^{1/2}\left(1+\frac{\Delta
r_1}{3l-r_2}\right)^{-1/2}\times
\\ \nl && F_D^{(6)}\left(1/2;-3/2,-1/2,-1/2,-1/2,-1/2,1/2;1;\right.
\\ \nl &&\left.\frac{1}{1+\frac{2l}{\Delta
r_1}},\frac{1}{1+\frac{4l}{\Delta
r_1}},\frac{1}{1+\frac{6l}{\Delta
r_1}},\frac{1}{1+\frac{3l-u+}{\Delta
r_1}},\frac{1}{1+\frac{3l-u-}{\Delta
r_1}}\frac{1}{1+\frac{3l-r_2}{\Delta r_1}}\right),\ea \ba\nl
&&K^D_{B2}=
2\pi^2T_{D2}\left[\frac{l^3\left(\hat{\Lambda}_{1+}^2+\hat{\Lambda}_{1-}^2\right)(3l-u_+)(3l-u_-)}
{3^3\left(\hat{\Lambda}_+^2+\hat{\Lambda}_-^2+4\hat{\Lambda}_D^2/3\right)\left(3l-r_2\right)}\right]^{1/2}\times
\\ \nl &&\Delta r_1\left(1+\frac{\Delta
r_1}{2l}\right)^{1/2}\left(1+\frac{\Delta
r_1}{4l}\right)^{3/2}\left(1+\frac{\Delta r_1}{6l}\right)^{-1/2}\times \\
\nl &&\left(1+\frac{\Delta r_1}{3l-u_+}\right)^{1/2}
\left(1+\frac{\Delta r_1}{3l-u_-}\right)^{1/2}\left(1+\frac{\Delta
r_1}{3l-r_2}\right)^{-1/2}\times
\\ \nl && F_D^{(6)}\left(1/2;-1/2,-3/2,1/2,-1/2,-1/2,1/2;2;\right.
\\ \nl &&\left.\frac{1}{1+\frac{2l}{\Delta
r_1}},\frac{1}{1+\frac{4l}{\Delta
r_1}},\frac{1}{1+\frac{6l}{\Delta
r_1}},\frac{1}{1+\frac{3l-u+}{\Delta
r_1}},\frac{1}{1+\frac{3l-u-}{\Delta
r_1}}\frac{1}{1+\frac{3l-r_2}{\Delta r_1}}\right),\ea \ba\nl
&&K^D_{D2}=
4\pi^2T_{D2}\left[\frac{l^5\left(\hat{\Lambda}_{1+}^2+\hat{\Lambda}_{1-}^2\right)(3l-u_+)(3l-u_-)}
{3\left(\hat{\Lambda}_+^2+\hat{\Lambda}_-^2+4\hat{\Lambda}_D^2/3\right)\left(3l-r_2\right)}\right]^{1/2}\times
\\ \nl &&\left(1+\frac{\Delta
r_1}{2l}\right)^{1/2}\left(1+\frac{\Delta
r_1}{3l}\right)^{2}\left(1+\frac{\Delta
r_1}{4l}\right)^{1/2}\left(1+\frac{\Delta r_1}{6l}\right)^{-1/2}\times \\
\nl &&\left(1+\frac{\Delta r_1}{3l-u_+}\right)^{1/2}
\left(1+\frac{\Delta r_1}{3l-u_-}\right)^{1/2}\left(1+\frac{\Delta
r_1}{3l-r_2}\right)^{-1/2}\times
\\ \nl && F_D^{(7)}\left(1/2;-1/2,-2,-1/2,1/2,-1/2,-1/2,1/2;1;\right.
\\ \nl &&\left.\frac{1}{1+\frac{2l}{\Delta
r_1}},\frac{1}{1+\frac{3l}{\Delta
r_1}},\frac{1}{1+\frac{4l}{\Delta
r_1}},\frac{1}{1+\frac{6l}{\Delta
r_1}},\frac{1}{1+\frac{3l-u+}{\Delta
r_1}},\frac{1}{1+\frac{3l-u-}{\Delta
r_1}}\frac{1}{1+\frac{3l-r_2}{\Delta r_1}}\right).\ea

Taking the semiclassical limit in the above expressions for $E$,
$P_I$ and $P_\phi$, which in the case under consideration
corresponds to \ba\nl r_{1,2}\to \pm
2\sqrt{\frac{3v_0^2}{\hat{\Lambda}_+^2+\hat{\Lambda}_-^2+4\hat{\Lambda}_D^2/3}}\to\infty,\ea
we receive that the energy depends on $P_I$ and $P_\phi$ as
follows \ba\label{ED6} E^2=\mathbf{P}^2+3^{7/3}\left(\frac{2\pi
T_{D2}
\Lambda_1^{\phi}\sin\theta^0}{4-\sin^2\theta^0}\right)^{2/3}P_{\phi}^{4/3}.\ea
This is another generalization of the energy-charge relation given
in (\ref{ED2}).

\setcounter{equation}{0}
\section{Comments and conclusions}
In this paper, we considered rotating strings and D2-branes on
type IIA background, which arises as dimensional reduction of
M-theory on manifold of $G_2$ holonomy, dual to $\mathcal{N}=1$
gauge theory in four dimensions. We obtained exact solutions and
explicit expressions for the energy and other momenta (charges),
which are conserved due to the presence of background isometries.
They were given in terms of the hypergeometric functions of many
variables $F_D^{(n)}(a;b_1,\ldots,b_n;c;z_1,\ldots,z_n)$, where
for the different cases considered, $n$ varies from one to seven.

We investigated the semiclassical limit of the conserved
quantities and received different types of relations between them.
Our aim was to check if strings and D2-branes rotating in this ten
dimensional type IIA background, can reproduce the energy-charge
relations obtained in \cite{27} and \cite{B0306} for rotating
M2-branes on $G_2$ manifolds. We found that the rotating strings
can reproduce only one type of semiclassical behavior, exhibited
by rotating M2-branes. Our results are the following \ba\nl
&&E^2=\mathbf{P}^2+2\pi T
\left(6r_0\right)^{1/2}\left(P_{\theta_1}^2+P_{\theta_2}^2\right)^{1/2},\\
\nl &&E^2=\mathbf{P}^2+2\pi T
\left(6r_0\right)^{1/2}\left(P_{\theta}^2+
\frac{3P_{\phi}^2}{3-\cos^2\theta_2^0}\right)^{1/2},\\ \nl
&&E^2=\mathbf{P}^2+2\pi T (6r_0)^{1/2}\times\\
\nl &&\left[\frac{\left(3-\cos^2\theta_2^0\right)P_{\phi_1}^2+
\left(3-\cos^2\theta_1^0\right)P_{\phi_2}^2
-4P_{\phi_1}P_{\phi_2}\cos\theta_1^0\cos\theta_2^0}
{3-\cos^2\theta_1^0-
\cos^2\theta_2^0-\cos^2\theta_1^0\cos^2\theta_2^0}\right]^{1/2}.\ea
These equalities are generalizations of the $E\sim K^{1/2}$
behavior and correspond to the following M2-brane energy-charge
relation \cite{B0306} \ba\label{M2}
&&\left\{E^2\left(E^2-\mathbf{P}^2\right) - (2\pi^2 T_{M2}
l_{11}^3)^2\left\{\left(\mathbf{\Lambda}_1\times\mathbf{\Lambda}_2\right)^2
E^2 -
\left[\left(\mathbf{\Lambda}_1\times\mathbf{\Lambda}_2\right)\times\mathbf{P}\right]^2\right\}\right\}^2
\\ \nl &&-6(2\pi^2 T_{M2} l_{11}^3)^2E^2\left[\mathbf{\Lambda}_1^2 E^2 -
\left(\mathbf{\Lambda}_1.\mathbf{P}\right)^2\right]\left(P^2_{\theta}+P^2_{\tilde{\theta}}\right) = 0.\ea

We also showed that the rotating D2-branes reproduce two types of
the semiclassical energy-charge relations known for membranes in
M-theory. The first type is represented by \ba\nl
&&\frac{E^2\left(E^2-\mathbf{P}^2\right)^2}{\mathbf{\Lambda}_1^2
E^2 - \left(\mathbf{\Lambda}_1.\mathbf{P}\right)^2}
=\frac{2^3}{3^5}(\pi^2 T_{D2})^2
\left(P^2_{\theta_1}+P^2_{\theta_2}\right),\\
\nl &&\frac{E^2\left(
E^2-\mathbf{P}^2\right)^2}{\mathbf{\Lambda}_1^2 E^2
-\left(\mathbf{\Lambda}_1.\mathbf{P}\right)^2}
=\frac{2^3}{3^5}(\pi^2 T_{D2})^2
\left(P^2_{\theta}+\frac{3P_{\phi}^2}{3-\cos^2\theta_2^0}\right),
\\ \nl &&\frac{E^2\left(
E^2-\mathbf{P}^2\right)^2}{\mathbf{\Lambda}_1^2 E^2
-\left(\mathbf{\Lambda}_1.\mathbf{P}\right)^2} =\\ \nl
&&\frac{2^3}{3^5}(\pi^2
T_{D2})^2\frac{\left(3-\cos^2\theta_2^0\right)P_{\phi_1}^2+
\left(3-\cos^2\theta_1^0\right)P_{\phi_2}^2
-4P_{\phi_1}P_{\phi_2}\cos\theta_1^0\cos\theta_2^0}
{3-\cos^2\theta_1^0-
\cos^2\theta_2^0-\cos^2\theta_1^0\cos^2\theta_2^0}.\ea These are
generalizations of the dependence $E\sim K^{1/2}$ and correspond
to (\ref{M2}). For the second type, we received the equalities
\ba\nl &&E^2=\mathbf{P}^2+3^{5/3}(2\pi T_{D2}
\Lambda_1^{\theta})^{2/3}P_{\theta}^{4/3},\\ \nl
&&E^2=\mathbf{P}^2+\frac{3^{7/3}}{2^{1/3}}\left(\frac{\pi T_{D2}
\Lambda_1^{\theta}}{3-\cos^2\theta_2^0}\right)^{2/3}P_{\phi}^{4/3},\\
\nl &&E^2=\mathbf{P}^2+3^{7/3}\left(\frac{2\pi T_{D2}
\Lambda_1^{\phi}\sin\theta^0}{4-\sin^2\theta^0}\right)^{2/3}P_{\phi}^{4/3},\ea
which are generalizations of the dependence $E\sim K^{2/3}$ and
correspond to \cite{B0306} \ba\nl E^2=\mathbf{P}^2+3^{5/3}(2\pi
T_{M2} l_{11}^3\Lambda_1^6)^{2/3}P_{\theta}^{4/3}.\ea

We were not able to obtain the other three types of semiclassical
behavior discovered in \cite{B0306} for M2-branes \ba\nl &&E^2=
\mathbf{P}^2 + \frac{9}{2l^2}P_+^2 - (6\pi^2T_{M2}
l_{11}^3\Lambda_1^-)^{2/3}P_+^{4/3},\\ \nl
&&\left\{E^2\left[E^2-\mathbf{P}^2-(3/l)^2P_+^2\right]-(2\pi^2T_{M2}
l_{11}^3)^2\left\{\left(\mathbf{\Lambda}_1\times\mathbf{\Lambda}_2\right)^2
E^2 -
\left[\left(\mathbf{\Lambda}_1\times\mathbf{\Lambda}_2\right)\times\mathbf{P}\right]^2\right\}\right\}^2
\\ \nl &&-2^7(3\pi T_{M2} l_{11}^3)^2E^2\left[\mathbf{\Lambda}_1^2 E^2 -
\left(\mathbf{\Lambda}_1.\mathbf{P}\right)^2\right]P_{+}^2 = 0,\\
\nl &&\left\{E^2\left[E^2-\mathbf{P}^2-(3/2l)^2 P_+^2\right] -
(2\pi^2 T_{M2}
l_{11}^3)^2\left\{\left(\mathbf{\Lambda}_1\times\mathbf{\Lambda}_2\right)^2
E^2 -
\left[\left(\mathbf{\Lambda}_1\times\mathbf{\Lambda}_2\right)\times\mathbf{P}\right]^2\right\}\right\}^2
\\ \nl &&-(6\pi^2 T_{M2} l_{11}^3)^2E^2\left[\mathbf{\Lambda}_1^2 E^2 -
\left(\mathbf{\Lambda}_1.\mathbf{P}\right)^2\right]P_{-}^2 = 0,\ea
which generalize the relations \ba\nl E-K\sim K^{1/3},\h E-K\sim
const,\h E\sim K_1 + const \frac{K_2}{K_1}.\ea One reason is that
after the dimensional reduction from eleven to ten dimensions, the
term in the background metric proportional to $C^2(r)$ disappears
(compare (\ref{10db}) with (\ref{G2-g})). Besides, we considered
very restricted class of solutions, depending only on the radial
background coordinate. However, these are just kind of technical
reasons. To our opinion, the physical cause behind is that other
types of M2-brane's semiclassical behavior should be reproduced in
ten dimensions by more complex non-perturbative states like bound
states of fundamental strings and D-branes. Support for this
conjecture are the results obtained in \cite{BRR04}, where such
relation has been found for flat space-time. More precisely,
starting with rotating membranes solutions in flat eleven
dimensions, and compactifying on a circle and on a torus, the
authors of \cite{BRR04} have been able to identify
non-perturbative states of type IIA and type IIB superstring
theory, which represent spinning bound states of D-branes and
fundamenal strings.

We note that in considering the semiclassical limit (large
charges), we take into account only the leading terms in the
expressions for the conserved quantities. However, there is no
problem to include the higher order terms. An example is given in
(\ref{pc3G'h}).

For comparison, we now give two known results about the
energy-charge relations, obtained in the semiclassical limit, for
strings moving in other curved type IIA backgrounds\footnote{See
also \cite{S03},where spinning and rotating closed string
solutions in $AdS_5 \times T^{1,1}$ background have been found,
and has been shown how these solutions can be mapped onto rotating
closed strings embedded in configurations of intersecting branes
in type IIA string theory.}.

Rotating strings in a warped $AdS_6\times S^4$ geometry have been
considered in \cite{CLP04}. The warped $AdS_6\times S^4$ is vacuum
solution of the massive type IIA supergravity, which is expected
to be dual to an $\mathcal{N}=2$, $D=5$ super-conformal Yang-Mills
theory. For large conserved charges, the following relation
between them has been found \ba\nl E- \frac{3}{2}J=c_1 +
\frac{c_2}{J^5}+\ldots.\ea At the leading order, this relation is
of the type $E-K\sim const$, and is reproduced by one of the
M2-brane configurations described above, but not by the strings
and D2-branes considered here.

Pulsating strings in the same warped $AdS_6\times S^4$ background
have been semiclassically quantized in \cite{BDR04} with the
result \ba\nl E^2=(J+7/3)(J+4) + \mbox{quantum corrections},\ea
which in the leading order gives the $E-K\sim const$ behavior once
again.

It seems to us that an interesting task, which deserves to be
investigated, is the semiclassical behavior of the strings and
D2-branes in the $\gamma$-deformed \cite{LM05} background
(\ref{10db}), in order to see the difference with the results
obtained here, and to estimate the role of the Kaluza-Klein modes,
following the idea developed in \cite{GN05}, and applied for
semiclassical strings in \cite{BDR11216}. This problem is under
investigation and we hope to report about some progress soon.

\vspace*{.5cm}

{\bf Acknowledgments} \vspace*{.2cm}

This work is supported by NSFB grant under contract $\Phi1412/04$.

\appendix
\setcounter{equation}{0}
\section{Hypergeometric functions $F_{D}^{(n)}$}

Here, we give some properties of the hypergeometric functions of many variables
$F_{D}^{(n)}$ used in our calculations. By definition \cite{PBM-III}, for $|z_j|<1$,
\ba\nl F_D^{(n)}(a;b_1,\ldots,b_n;c;z_1,\ldots,z_n)=
\sum_{k_1,\ldots,k_n=0}^{\infty}\frac{(a)_{k_1+\ldots+k_n}(b_1)_{k_1}\ldots(b_n)_{k_n}}
{(c)_{k_1+\ldots+k_n}}\frac{z_1^{k_1}\ldots z_n^{k_n}}{k_1!\ldots k_n!},\ea
where
\ba\nl (a)_k = \frac{\Gamma(a+k)}{\Gamma(a)},\ea
and $\Gamma(z)$ is the Euler's $\Gamma$-function.
In particular, $F_D^{(1)}(a;b;c;z)= \mbox{}_2F_{1}(a,b;c;z)$ is the Gauss' hypergeometric function,
and $F_D^{(2)}(a;b_1,b_2;c;z_1,z_2)= F_{1}(a,b_1,b_2;c;z_1,z_2)$ is one of the hypergeometric
functions of two variables.
\ba\nl
{\bf 1.} &&F_D^{(n)}(a;b_1,\ldots,b_i,\ldots,b_j,\ldots,b_n;c;z_1,\ldots,z_i,\ldots,z_j,\ldots,z_n)=
\\ \nl &&F_D^{(n)}(a;b_1,\ldots,b_j,\ldots,b_i,\ldots,b_n;c;z_1,\ldots,z_j,\ldots,z_i,\ldots,z_n).
\\ \nl
{\bf 2.} &&F_D^{(n)}(a;b_1,\ldots,b_n;c;z_1,\ldots,z_n)=
\\ \nl &&\prod_{i=1}^{n}\left(1-z_i\right)^{-b_i}
F_D^{(n)}\left(c-a;b_1,\ldots,b_n;c;\frac{z_1}{z_1-1},\ldots,\frac{z_n}{z_n-1}\right).
\\ \nl
{\bf 3.} &&F_D^{(n)}(a;b_1,\ldots,b_{i-1},b_i,b_{i+1},\ldots,b_n;c;z_1,\ldots,z_{i-1},1,z_{i+1},\ldots,z_n)=
\\ \nl &&\frac{\Gamma(c)\Gamma(c-a-b_i)}{\Gamma(c-a)\Gamma(c-b_i)}
F_D^{(n-1)}(a;b_1,\ldots,b_{i-1},b_{i+1},\ldots,b_n;c-b_i;z_1,\ldots,z_{i-1},z_{i+1},\ldots,z_n).
\\ \nl
{\bf 4.} &&F_D^{(n)}(a;b_1,\ldots,b_{i-1},b_i,b_{i+1},\ldots,b_n;c;z_1,\ldots,z_{i-1},0,z_{i+1},\ldots,z_n)=
\\ \nl &&F_D^{(n-1)}(a;b_1,\ldots,b_{i-1},b_{i+1},\ldots,b_n;c;z_1,\ldots,z_{i-1},z_{i+1},\ldots,z_n).
\\ \nl
{\bf 5.} &&F_D^{(n)}(a;b_1,\ldots,b_{i-1},0,b_{i+1},\ldots,b_n;c;z_1,\ldots,z_{i-1},z_i,z_{i+1},\ldots,z_n)=
\\ \nl &&F_D^{(n-1)}(a;b_1,\ldots,b_{i-1},b_{i+1},\ldots,b_n;c;z_1,\ldots,z_{i-1},z_{i+1},\ldots,z_n).
\\ \nl
{\bf 6.} &&F_D^{(n)}(a;b_1,\ldots,b_i,\ldots,b_j,\ldots,b_n;c;z_1,\ldots,z_i,\ldots,z_i,\ldots,z_n)=
\\ \nl &&F_D^{(n-1)}(a;b_1,\ldots,b_i+b_j,\ldots,b_n;c;z_1,\ldots,z_i,\ldots,z_n).
\\ \nl
{\bf 7.} &&F_D^{(2n+1)}(a;a-c+1,b_2,b_2,\ldots,b_{2n},b_{2n};c;-1,z_2,-z_2\ldots,z_{2n},-z_{2n})=
\\ \nl &&\frac{\Gamma(a/2)\Gamma(c)}{2\Gamma(a)\Gamma(c-a/2)}
F_D^{(n)}(a/2;b_2,\ldots,b_{2n};c-a/2;z_2^2,\ldots,z_{2n}^2).
\\ \nl
{\bf 8.} &&F_D^{(2n+1)}\left(c-a;a-c+1,b_2,b_2,\ldots,b_{2n},b_{2n};c;\right. \\ \nl  &&\left.1/2,-\frac{z_2}{1-z_2},\frac{z_2}{1+z_2},\ldots,-\frac{z_{2n}}{1-z_{2n}},\frac{z_{2n}}{1+z_{2n}}\right)=
\\ \nl &&\frac{\Gamma(a/2)\Gamma(c)}{2^{c-a}\Gamma(a)\Gamma(c-a/2)}
F_D^{(n)}\left(c-a;b_2,\ldots,b_{2n};c-a/2;-\frac{z_2^2}{1-z_2^2},\ldots,-\frac{z_{2n}^2}{1-z_{2n}^2}\right).
\\ \nl
{\bf 9.} &&F_D^{(2)}\left(a;b,b;c;z,-z\right)=
\mbox{}_3F_{2}\left(\matrix{a/2,(a+1)/2,b \\ \nl c/2,(c+1)/2;z^2}\right).\cr\ea


\end{document}